\documentclass{emulateapj}
\usepackage{epsfig}
\usepackage{natbib}

\def\deg{\ifmmode {^\circ}\else {$^\circ$}\fi} 
\def\arcmin{\ifmmode^{\prime}\else $^{\prime}$\fi} 
\def\arcsec{\ifmmode^{\prime \prime}\else $^{\prime \prime}$\fi} 
\def\msun{\ifmmode {\rm M_{\odot}}\else $\rm M_{\odot}$\fi} 
\def\X{$\times~$}
\def\Kalpha{K$\alpha$~}
\def\ergcm2s{ergs~cm$^{-2}$~s$^{-1}$}  
\def\cm2s{cm$^{-2}$~s$^{-1}$}  
\def\redchi2{$\chi_\nu^2$}
\def\1543{4U~1543$-$47}

\shorttitle{ 2002 Outburst of \1543}
\shortauthors{Park et al.}

\begin{document}

\title{Spectral and Timing Evolution of the Black Hole \\ X--ray Nova 4U~1543$-$47 During its 2002 Outburst}

\author{S.~Q.~Park\altaffilmark{1}, 
        J.~M.~Miller\altaffilmark{1,2},
        J.~E.~McClintock\altaffilmark{1}, 
        R.~A.~Remillard\altaffilmark{3}, 
        J.~A.~Orosz\altaffilmark{4}, 
        C.~R.~Shrader\altaffilmark{5},
        R.~W.~Hunstead\altaffilmark{6},
        D.~Campbell--Wilson \altaffilmark{6},
        C.~H.~Ishwara--Chandra\altaffilmark{7},
        A.~P.~Rao\altaffilmark{7},
        M.~P.~Rupen\altaffilmark{8}}

\altaffiltext{1}{Harvard--Smithsonian Center for Astrophysics, 60 Garden Street, Cambridge, MA 02138; spark@cfa.harvard.edu, jmmiller@cfa.harvard.edu, jem@cfa.harvard.edu} 
\altaffiltext{2}{NSF Astronomy and Astrophysics Fellow} 
\altaffiltext{3}{Center for Space Research, MIT, Cambridge, MA 02139; rr@space.mit.edu} 
\altaffiltext{4}{Department of Astronomy, San Diego State University, 5500 Campanile Drive, San Diego, CA 92182--1221; orosz@zwartgat.sdsu.edu} 
\altaffiltext{5}{Laboratory for High--Energy Astrophysics, NASA Goddard Space Flight Center, Greenbelt, MD 20771; shrader@grossc.gsfc.nasa.gov}
\altaffiltext{6}{School of Physics, University of Sydney, NSW 2006, Australia; rwh@physics.usyd.edu.au, dcw@physics.usyd.edu.au}
\altaffiltext{7}{National Center for Radio Astrophysics, TIFR, Post Bag 3, Ganeshkhind, Pune 411 007, India; pramesh@ncra.tifr.res.in, ishwar@ncra.tifr.res.in}
\altaffiltext{8}{National Radio Astronomy Observatory, Socorro, NM 87801; mrupen@zia.aoc.nrao.edu}

\begin{abstract} 
We present an X--ray spectral and timing analysis of \1543 during its
2002 outburst based on 49 pointed observations obtained using the {\it
Rossi X--ray Timing Explorer} {\it (RXTE)}.  The outburst reached a
peak intensity of 4.2 Crab in the 2--12 keV band and declined by a
factor of 32 throughout the month--long observation.  A $21.9 \pm 0.6$
mJy radio flare was detected at 1026.75~MHz two days before the X--ray
maximum; the radio source was also detected late in the outburst,
after the X--ray source entered the {\it hard} state.  The X--ray
light curve exhibits the classic shape of a rapid rise and an
exponential decay.  The spectrum is soft and dominated by emission
from the accretion disk.  The continuum is fit with a multicolor disk
blackbody ($kT_{max} = 1.04$ keV) and a power--law ($\Gamma \sim
2.7$).  Midway through the decay phase, a strong low--frequency QPO
($\nu$ = 7.3--8.1~Hz) was present for several days.  The spectra
feature a broad Fe \Kalpha line that is asymmetric, suggesting that
the line is due to  relativistic broadening rather than
Comptonization.  Relativistic Laor  models provide much better fits to
the line than non--relativistic  Gaussian models, particularly near
the beginning and end of our observations.  The line fits yield
estimates for the inner disk radius  that are within 6 $R_g$; this
result and additional evidence indicates  that this black hole may
have a non--zero angular momentum.
\end{abstract} 

\keywords{ 
accretion, accretion disks ---
binaries: close ---
black hole physics ---
stars: individual (IL Lupi, \1543) --- 
X--rays: stars 
}

\maketitle 

\section{INTRODUCTION} 

The recurrent X--ray nova \1543 was observed in outburst in 1971,
1983, 1992 and 2002 (Matilsky et al.\ 1972; Kitamoto et al.\ 1984;
Harmon et al.\ 1992; this work).  During the recent outburst, the peak
X--ray intensity was 4.2 times that of the Crab Nebula (2--12 keV).
The peak intensities during the three previous outbursts were
comparable (Tanaka \& Lewin 1995).  The X--ray spectrum indicates that
the primary is a black hole candidate: the 1--10 keV spectrum is
invariably soft when the source is bright, and in 1992 a power--law
component was observed with an intensity of $\approx0.3$~Crab in an
energy band extending from 120 to 230 keV (Harmon et al.\ 1992; Tanaka
\& Lewin 1995).

The optical counterpart of \1543, IL Lupi, was discovered by Pederson
(1983).  The secondary star has a  relatively large mass and early
spectral type, A2V (Chevalier \& Ilovaisky 1992; McClintock \&
Remillard 2003).  Because the secondary is relatively luminous, the
optical counterpart brightens by only $\approx1.8$~mag in outburst
(van Paradijs \& McClintock 1995), although the X--ray source
brightens by an enormous factor, $>2~\times~10^7$ (Garcia et al.\
2001), which is typical for an X--ray nova (McClintock \&
Remillard 2003).  In the decade--long intervals of quiescence, the
source intensity is $\lesssim$ 0.1 $\mu$Crab.   In this inert state,
IL Lupi has been the subject of detailed dynamical studies by Orosz et
al.\ (1998, 2003), who have derived the following primary data for
this 26.8--hr binary: a black hole mass of $M_{1} = 9.4\pm2.0$~\msun,
a secondary star mass of $M_{2} = 2.7\pm1.0$~\msun, and a distance of
$D = 7.5\pm1.0$~kpc (where the distance error includes an uncertainty
in the reddening of $E_{B-V}=0.1$~mag).  The large
value of $M_{1}$ confirms that the compact primary is a black hole
(Rhoades \& Ruffini 1974).  Of special importance for the present
work, which focuses on the Fe \Kalpha line, \1543 has a very low
orbital inclination: $i = 20.7\pm1.0^{\rm o}$ (Orosz et al. 2003).

During the 1983 outburst, a broad (FWHM~$\sim2.7$~keV) line centered
at 5.9~keV was observed using the Gas Scintillation Proportional
Counter (GSPC) aboard {\it EXOSAT} (van der Woerd, White \& Kahn 1989).  The
higher resolution of the GSPC (relative to our study) makes this a
pertinent result, which we return to in \S4.2.  Van der Woerd et al.\
interpreted this feature as a redshifted and broadened Fe \Kalpha line
formed by Compton scattering and/or by Doppler and relativistic
effects.  More recently, a number of very {\it asymmetric} Fe \Kalpha
line profiles with extended red wings have been observed in both
Seyfert galaxies and black hole binaries (e.g., Tanaka \& Lewin 1995;
Miller et al.\ 2002).  Relativistic beaming and gravitational
redshifts in the inner disk region can serve to create such asymmetric
profiles.  Such broad Fe \Kalpha lines are thought to be generated by
fluorescence through the irradiation of the accretion disk by a source
of hard X--rays, presumably a Comptonizing corona.
(For recent reviews, see Fabian et al.\ 2000 and Reynolds \& Nowak
2003.)

In the present work, we report on a 36--day campaign of 49 pointed 
observations made using the {\it Rossi X--ray Timing Explorer} {\it
(RXTE)}  Proportional Counter Array (PCA).  We present the results of 
a spectral analysis that focuses on the behavior of the often intense,
broad Fe \Kalpha line.  We also present energy and power spectra for 
selected pointed observations.  In \S2 we discuss the observations and
data reduction.  The results of the spectral fitting are presented in 
\S3, and a discussion of the Fe line and low frequency QPOs follows in
\S4.  We conclude in \S5 with a summary of our results. 

\section{OBSERVATIONS AND DATA REDUCTION} 

We observed 4U 1543$-$47 with {\it RXTE} on 49 occasions between
Modified Julian Date (MJD = Julian Date $-$ 2,400,000.5)  52442--52477
(2002 June 17 -- 2002 July 22).  These data correspond to all of the
{\it RXTE} pointed observations made under  programs 70133 and 70132,
excluding a pair of observations made on  2002 July 1 that yielded
problems with the archival data files.  For  four of the longest
observations (Table 1: 0617, 0619, 0620, and 0628), we divided the
exposures into two observing intervals.  A list of the observation
midpoint times and durations is given in Table  1.  Owing to the
extremely soft nature of this source, we did not include High Energy
X--ray Timing Experiment (HEXTE) data in our analysis.

\subsection{Spectroscopy} 

At the time of our data analyses, PCU--2 was the best--calibrated PCU
in the {\it RXTE}/PCA (based on fits to observations of the Crab,
which is taken  to be a simple power--law in the {\it RXTE} band).  As
we wish to examine sometimes subtle Fe \Kalpha emission lines in the
spectra of  4U 1543$-$47, we have therefore restricted our spectral
analysis to include only data from PCU--2.

Data reduction tools from HEASOFT version 5.2 were used to screen the
event files and spectra.  Data were taken in the ``Standard 2 mode'',
which provides coverage of the full PCA bandpass (2--60~keV) every 16
seconds.  Data from all Xe gas layers of PCU--2 were added to make the
spectra.  Background spectra were made using the tool ``pcabackest''
using the latest ``bright source'' background model.  Background
spectra were subtracted from the total spectra using the tool
``mathpha''.  Redistribution matrix files (rmfs) and ancillary
response files (arfs) were generated for each PCU layer and combined
into a single response file using the tool ``pcarsp.''

It is well--known that fits to PCA spectra of the Crab nebula reveal
residuals as large as 1\%, and it has become customary to add
systematic errors of 1\% to all PCU energy channels (see, e.g.,
Sobczak et al.\ 2000).  We therefore added systematic errors of 1\% to
all PCU--2 energy channels using the tool ``grppha.''   The spectral
fits yielded no signs of residuals attributable to absorption from the
Xe L3 edge at 4.79~keV (see, e.g., Miller et al.\ 2001; and Ibrahim,
Swank, and Parke 2003), and therefore we did not include such an edge
in our spectral model.  Finally, strong deviations which cannot be
accounted for by any plausible source spectral feature are found below
3~keV, and the  spectrum becomes background--dominated and the
calibration less certain above 25~keV.  We therefore restricted our
spectral analysis to the 2.9--25.0~keV band, which is customary for
analysis of PCA spectra obtained after the gain change of 1999 March.

\subsection{Timing} 

X--ray power density spectra (PDS) and the search for quasi--periodic 
oscillations (QPO) follow the descriptions given by Remillard et
al.\ (2002).  PDS in the frequency range 4~mHz to 4~kHz were computed 
for data in the energy range 2--30~keV for all of the observations
listed in Table 1, with the exception of observations 10 and 11, where 
the data from fast PCA timing modes was not successfully telemetered.
The data modes used for program 70133 provided opportunities to 
compute additional PDS to 4 kHz over any combination of energy bands:
2--6, 6--15, and 15--30~keV.

\section{ANALYSIS AND RESULTS} 

The All Sky Monitor (ASM) total light curve from 2--12~keV and the ASM
(5--12~keV)/(3--5~keV) hardness ratio (HR2) are plotted in Figure 1.
The spectrum has a hardness ratio of $\approx0.5$ throughout most of
the outburst,  suggesting that the spectrum is soft (McClintock \&
Remillard 2003) and that thermal emission from an optically thick
accretion disk is  the dominant component of the spectrum (Shakura \&
Sunyaev 1973).  The light curve exhibits the classic shape of a rapid
rise and an  exponential decay (Chen, Shrader, \& Livio 1997;
McClintock \& Remillard 2003), with the source reaching a maximum
intensity of 4.2  Crab at 2--12~keV near MJD 52445 (2002 June 20).
The spectrum was initially hard during the rise to maximum.  The
source then transitioned to the {\it  thermal dominant} (TD) state
(McClintock \& Remillard 2003), which is often referred to as the {\it
high soft} state, where it remained for  the next $\sim30$ days.  For
several days during the decay phase, the source exhibited
low--frequency QPOs and remained in the {\it steep  power law} state
(see \S3.2), which is similar to the {\it intermediate} or {\it very
high} state of black hole binaries.  Finally, on MJD 52475 (2002 July
20) it entered the {\it hard} state at an intensity of
$\approx0.1$~Crab.   The ASM light  curve can be compared with the PCA
light curve in Figure~2, which is based on 49 pointed observations.
The figure shows the total 2.9--25.0~keV unabsorbed flux over the
observation period, along with the disk, power--law, and line
components of the total flux.

All spectral analyses were performed using XSPEC version 11.2 (Arnaud
1996).  Unless a source spectrum is strongly absorbed by the
interstellar medium, the 2.9--25.0~keV band is not well--suited to
measuring the equivalent neutral hydrogen column density.  Attempts to
determine this parameter via the \1543 spectra using the
``phabs'' model consistently yielded error bars which included zero at
90\% confidence.  We therefore fixed this parameter to $N_{H} = 4.0
\times 10^{21}~{\rm cm}^{-2}$ (Dickey \& Lockman 1990, as implemented
in the $N_{H}$ tool on the HEASARC website).  The transmission is 93\%
even at 2.9~keV, and therefore interstellar absorption is relatively
unimportant. 

In addition to interstellar absorption, the spectra were fit to a
model that incorporates several other elements, including a multicolor
blackbody accretion disk component (Mitsuda et al.\ 1984; Makishima et
al.\ 1986) and a power--law component.  The fits returned the
following continuum parameters: the color temperature at the inner
edge of the disk in keV ($T_{col}$), the disk normalization parameter
($K$), the power--law photon index ($\Gamma$), and the power--law
normalization factor.  The disk normalization parameter $K$ allows for
the computation of an approximate inner disk radius ($R_{col}$):
\begin{equation} 
K = \left(\frac{R_{col}/km}{D/10~kpc}\right)^2 cos~i
\end{equation}
\noindent where $D$ = 7.5~kpc is the distance to the source and $i$ =
21\deg~is the inclination angle of the disk.  Throughout this work we
express the radius in units of the gravitational radius by dividing by
$R_g \equiv GM/c^2$, where $M$ = 9.4~\msun~is the black hole mass.
Thus, for \1543, $r_{col} = R_{col}/R_{g} = 0.056 K^{1/2}$ (in units of $R_g$).  We
use the subscript ``col'' as a reminder that this estimate of the
inner disk radius is based on a measurement of the apparent color temperature
that is derived from a model that does not include effects
of general relativity or electron scattering (see below).  The values
of the four continuum parameters over the course of the observations
are summarized in Table 1 and plotted in Figure 3.  

The thermal component of the emergent spectrum can be approximated as
a diluted blackbody if electron scattering dominates over absorption
as a source of opacity in the disk and if the X--ray spectrum is
affected by Comptonization (Shakura \& Sunyaev 1973; Ebisawa et al.\
1994): 
\begin{equation} 
I(E) = \left(\frac{1}{f^4}\right) B(T_{col}, E),
\end{equation} 
\noindent where B is the Planck function and $f$ is the spectral
hardening factor.  The values for $T_{col}$ and $r_{col}$ obtained
from the fitting procedure systematically underestimate the actual
disk radius, and the hardening factor $f$ is generally needed to
correct the results of the multicolor disk blackbody model, where the
effective temperature and the effective radius are given by
$T_{eff}=T_{col}/f$ and $r_{eff} = f^2r_{col}$ (Shimura \& Takahara
1995, see also Titarchuk \& Shrader 2002).  Merloni, Fabian, \& Ross
(2000) have presented two important results concerning corrections to
the multicolor disk blackbody model.  First, at low mass accretion
rates and/or when a high fraction of the power is dissipated in the
corona, simple correction factors  do not allow physical disk
parameters to be traced accurately.  Second, when the mass accretion
rate is high and/or a small fraction of the power is dissipated in the
corona, the multicolor disk blackbody model gives results in which
$r_{col}$ underestimates the physical size of the inner disk by a
constant factor (less than 2) over a broad range of conditions.
A strong majority of the spectra we have obtained from 
\1543 indicate that the second scenario holds, and we have
therefore not applied a correction factor.  We note that there are two
periods where the power--law becomes relatively more important (in the
middle of the campaign and at the end); however, for consistency we
have not applied any correction to these spectral results.  The source
also has a very low inclination (21\deg), which further lessens the
need for a hardening correction,  since a low inclination angle
generally implies a low optical depth to the source.

A broad smeared absorption edge (``smedge'' in XSPEC) near 7.5~keV was
included to approximately model the reflected continuum.  Following
Sobczak et al.\ (1999), we fixed the width at 7~keV, although we found
that varying the width between 5--10~keV had little effect on our results.
The returned parameters are the Fe edge energy in keV and the
optical depth in the Fe line, $\tau$ (see Table 1).

The Fe \Kalpha feature was fit with a Laor model (Laor 1991), which
assumes a Kerr black hole and accounts for relativistic effects, with
the line energy fixed between 6.4~keV (neutral Fe) and 7~keV
(H--like Fe).  The following parameters of the Laor model were fixed:
the power--law  index of the radial dependence of emissivity at $q =
3$ (where $J \propto r^{-q}$), the outer radius at 400 gravitational
radii, and the inclination angle at 21\deg.  The fits to the Laor
model returned values for line energy in keV, inner radius $\rho$ in
units of $R_g$, and line normalization factor in units of photons
\cm2s.  Figure 4 shows the evolution of $\rho$ throughout our observation.

For comparison, the line was also fit with a Gaussian model (again
with the line energy bounded between 6.4--7~keV).  This line model
returned parameters for line energy in keV, line width in keV, and
line normalization in photons \cm2s.  The fits to the Gaussian model
yielded higher values of $\chi_{\nu}^2$ than fits to the Laor model,
particularly near the beginning and end of the data set.  Figure 5
shows the comparison between reduced chi--squared values for the Laor
model and the Gaussian model for 40 degrees of freedom.  Each of these
models had a total of 9 free parameters (4 continuum, 2 smedge, and 3
line).  The superiority of the fits obtained using the Laor model can
be clearly illustrated by dividing our data into three sets.  We find
that the first 11 observations have a mean reduced chi squared value
of 1.773 using the Gaussian model and 1.247 using the Laor  model.
The probabilities of exceeding these values of $\chi^2_{\nu}$ are $p =
0.0019$ and $p = 0.14$, respectively.  The last 11 observations have a
$\chi^2_{\nu} =$ 1.496 (Gaussian) and 0.917 (Laor), with $p = 0.23$
and 0.63, respectively.  The middle 27 observations have mean
$\chi^2_{\nu}$ of 0.904 (Gaussian) and 0.790 (Laor), with $p = 0.64$
and 0.83, respectively.

Figure 6 shows some sample spectra from the observations.  The first
pair of panels (a and b) show the ratio of the spectrum to the model
for the observation with the worst fit, which occurred near the peak
of the outburst.  Panel (a) shows the observation after being fit with
just the interstellar absorption, disk blackbody, and power--law
components.  Panel (b) shows the same observation after the Laor and
smedge components have been included.  The addition of these
components reduces the \redchi2 drastically from 8.98 to 1.63.  A more
representative observation before and after the Laor and smedge
components is shown in panels (c) and (d), respectively.  The \redchi2
in this example decreased from 4.01 in panel (c) to 0.94 in panel (d).
 
Tables 1 and 2 list the returned values for all of the model
parameters for fits to all 49 spectra. Unless otherwise noted, all
errors reported in this work are at the 90\% level of confidence.

\subsection{Evolution of the Spectral Parameters}

The color temperature of the inner disk increases through the first 7
observations, and then decreases nearly monotonically during the rest
of the observations (Fig.\ 3a).  The color temperature is a factor of
2 lower in the final observations than in the initial observations.
This trend in the color temperature closely traces the rise and fall
of the disk flux (Fig.\ 2a \& 3a).  It is interesting to note that
through the relatively smooth rise and long--term decline in color
temperature, the implied inner disk radius remains rather steady
(Fig.\ 3b).  Indeed, the inner disk extent as measured by both the disk
blackbody and Laor models does not appear to be correlated with the
inner disk temperature or the disk flux during the phase of the
outburst covered by our observations.  However, $r_{col}$ decreases
near MJD 52460 when the power--law becomes strong.  This is consistent
with the idea stated above that simple correction factors do not allow
physical disk parameters to be determined accurately when the
power--law component is strong.  It is important to note that the
values of $r_{col}$ are only approximate estimates, although the
relative changes between the values are significant (see Section 4.3).
We note that the 33\% dip seen in the color radius during observations 
18--20 (Table 1) may be primarily caused, in fact, by the simple effects
of scattering losses as we view the disk through an enhanced corona
that supplies the increased power--law flux during those same observations.

As the 2.9--25.0~keV emission is dominated by the disk, the total flux
more closely traces the disk flux than the power--law flux.   The
power--law is strongest in the middle of our set of observations  (MJD
52458--52464, approximately; Fig.\ 2c).  During this time, and again toward  the
end of our set of observations (after MJD 52470), the fraction of the
total flux contributed by the power--law exceeds 50\%.  Overall, the
power--law index is remarkably steady (Fig.\ 3c).  Moreover, the power--law index
and normalization do not appear to be correlated on long timescales.
However, on two occasions (near MJD 52444 and MJD 52471) the
power--law  normalization flared to higher values and on both
occasions the power--law index softened considerably (see Fig.\ 3).

In Figure 7, we have plotted the power--law flux versus line flux and
the disk component flux versus line flux.  It is clear that both the disk
and power--law fluxes are generally positively correlated with the
line flux.  The simplest picture of the accretion flow geometry around
a Galactic black hole at a high mass accretion rate is that an
accretion disk fuels a corona by the Compton--upscattering of disk
photons.  The corona in turn irradiates the disk and produces an
Fe~K$\alpha$ fluorescence emission line.  The mass accretion rate
through the disk should drive the disk, corona, and line flux on
timescales comparable to the viscous timescale in the disk.  The
correlations shown in Figure 7b are broadly consistent with this
simple picture.  However, the measured relationship between the
power--law flux and the line flux (Fig.\ 7a) unexpectedly deviates
from a simple correlation, showing two branches with offset
correlations between the two fluxes.  The group with higher line
emission relative to the power--law flux corresponds to observations
4--10 (see Table 2).  These data were obtained during the interval of
peak X--ray luminosity.  It is conceivable that the increased line
flux at such times is related to a higher ionization state near
the surface of the disk, which would increase the strength of the line
(Ross, Fabian \& Young, 1999).  This effect is not evident in Table 2;
however, the uncertainties are often as large as the expected line
shift, and so the results are not conclusive.  Alternatively, there
may be correlations between the overall luminosity and the geometry of
the power--law emitting region.  In this case, the times of highest
overall luminosity may be associated with a power--law emitting region
that more efficiently illuminates the disk.  This suggestion will be
further discussed in Section 4.3.

\subsection{Timing Results}

The average PDS over four time intervals are shown in Figure 8. Within
the indicated time intervals, the power spectra generally resemble the
displayed average; furthermore, there are only modest changes in the
X--ray spectral results within a given time interval (see Table 1).
Observations that lie in the gaps between these time intervals
(viz. MJD 52456--52458 and MJD 52474--52475) show characteristics of
transition, and  they are therefore excluded from the PDS averages.

The average PDS over MJD 52442--52455 shows a smooth power continuum
that scales approximately as $P_\nu \propto \nu^{-1}$ (see Fig.\ 8a).
During this time, as noted earlier, the disk displays a roughly
constant color temperature and radius, while the unabsorbed flux from
the disk ($f_{disk}$) exceeds that of the power--law component
($f_{pow}$). We find that throughout this time: $f_{disk} / (f_{disk}
+ f_{pow}) \ga 0.7$, for fluxes integrated over the range of the PCA
spectral fits (viz. 2.9--25.0~keV; see Fig.\ 2).  The dominance of the
disk is comparably evident if one chooses to integrate the flux over
wider keV limits. For example, using the bolometric disk flux and the
power--law flux at 1--25~keV, we find that the disk contributes more
than 80\% of the total flux (unabsorbed) throughout this time.  These
combined spectral and PDS characteristics are entirely consistent with
the {\it thermal--dominant} (TD) state of black hole binaries (aka
{\it high soft} state; McClintock \& Remillard 2003).

A transition out of the TD state is seen during MJD 52456--52458, and
weak QPOs appear in the PDS. On MJD 52457, a QPO  is seen ($6 \sigma$)
centered at 10.1~Hz with an rms amplitude $r \sim 1.0$\% and a
coherence parameter $Q = \nu / \Delta \nu_{FWHM} \sim 8$. Similar
results are seen on MJD 52458, with a QPO at $\nu = 10.0$~Hz, $r =
1.2$\%, and $Q \sim 5$. The QPO abruptly strengthens during MJD
52459--52461, and the average PDS is shown in Figure 8b. These three
observations yield $\nu =$ 7.3--8.1~Hz, $r =$ 5.7--8.4\%, and $Q
\sim$ 6--9. The QPO profiles show obvious harmonic features on MJD
52459 and 52461, and these are also seen in the average PDS for this
time interval.  The bumps on either side of the strongest peak appear
at frequencies $ 0.5 \nu $ and $ 2 \nu $, as seen in other sources
such as XTE~J1550$-$564 (Remillard et al.\ 2002). During the interval
of MJD 52459--52461, the spectral parameters shift toward smaller
color radius and a stronger flux contribution from the power--law
component: $f_{disk} / (f_{disk} + f_{pow}) \sim 0.45 $ for
2.9--25.0~keV (or 0.7--0.8 if one uses the bolometric disk flux and
the 1--25~keV power--law flux).   The QPO detection, the increased
strength of the power--law component, and the values of the photon
index (2.5--2.7; see Table 1) are all consistent with the {\it steep
power--law} (SPL) state of black hole binaries (McClintock \&
Remillard 2003).

After MJD 52461, the QPO abruptly disappears, as the PDS of MJD 52462
shows only a smooth continuum.  Meanwhile, the power--law flux rapidly
decreases, and the disk color radius rebounds to values consistent
with the MJD 52442--52455 results (Fig.\ 8a).  These characteristics
indicate a return to the TD state (Fig.\ 8c).  The TD state continues with a
gradually falling temperature, until MJD 52474, when the power--law
spectrum and PDS continuum power both begin to build again. During the
final days of our monitoring observations (MJD 52476--52477), a broad
power peak in seen in the PDS near 2~Hz and a weak QPO returns at 9.7
Hz (see Fig.\ 8d).  The photon index ($\Gamma = 2.4$) is lower at this
time, and this trend continues in subsequent days, reaching $\Gamma =
1.7$ on MJD 52487, when the source has essentially completed its
transition to the {\it hard} state (Kalemci et al.\ 2002;
McClintock \& Remillard 2003).

We searched for high--frequency QPOs (50--1000~Hz) in the PDS of each
observation, in each available energy band. No significant detections
were found.  In the energy range 6--30~keV, the statistical upper
limits ($4 \sigma$) for a QPO at 200~Hz (assuming $Q \sim 5$) are $r <
0.9$\% for the average PDS in the SPL state (MJD 52459--52461), and $r
< 0.4$\% for the TD state of MJD 52442--52455.  At 800 Hz, again
considering the average PDS shown in Fig.\ 8, the upper limits are
only slightly lower at 0.7\% for the SPL state and 0.3\% for the TD state.

The X--ray light curves (1 s bins) for \1543 always show the
random flickering that is characteristic of black hole binaries.
However, during the observation of 2002 July 6 (MJD 52461), which is
the last day that the source was in the SPL state, there are
hard dips suggestive of an accretion instability.  This light curve is
shown along with a PCA hardness ratio (HR) in Fig.\ 9.
The dips generally last 5--10 s, the dip rate is roughly one per
minute, and the dip minima have a constant value to within a few
percent.  We have also examined the spectral variations during the dips
using a different HR quantity that selects 6--15 keV versus 2--6 keV.
In this case there are almost no HR changes during the dips,
despite the improved statistics.  The latter result rules out the
possibility that the dips are associated with photoelectric absorption
due to intervening clouds.
 
The dips seen in \1543 are reminiscent of some of the patterned
variations seen in GRS~1915$+$105 (Belloni et al.\ 2000), which exhibits
an amazing repertoire of unstable light curves. However, the dips
in GRS~1915$+$105 with similar timescales (i.e.\ the $\gamma$ and
$\delta$ class variations) are spectrally ``soft'', unlike those in
\1543.  We further note that moderately unstable light curves
have also been seen in GRO~J1655$-$40 (Remillard et al. 1999), and
dipping behavior has been seen on several occasions from 4U~1630$-$47
(Tomsick, Lapshov, \& Kaaret 1998; Dieters et al. 2000).

\subsection{Radio Observations and Results}

Simultaneous radio observations were made during the X--ray outburst
using two radio telescopes.  The Molonglo Observatory Synthesis
Telescope (MOST) observed the source at 843~MHz with a 3~MHz bandwidth
in right--handed circular polarization; the synthesized beamwidth was
$58\arcsec \times 43\arcsec$ in position angle (PA) $0^{\circ}$.   The
Giant Metrewave Radio Telescope (GMRT), an aperture synthesis
instrument consisting of thirty parabolic dishes each 45~m in
diameter, observed the source in spectral line mode using a 16~MHz
bandwidth.  The GMRT observations were made at 1026.75~MHz (beamwidths
$\sim 9\arcsec \times 3\arcsec$), 1286.75~MHz ($6\arcsec \times
2\arcsec$), and 616.75~MHz ($12\arcsec \times 5\arcsec$), all with PA
near $0^{\circ}$.  The  observations at 616.75~MHz were made using
circularly polarized feeds, while the feeds at higher frequencies were
linearly polarized; in all cases both independent polarizations were
recorded.  The errors given for the MOST detections represent the
root-mean-square (rms) noise level measured near the source, while the
errors for the GMRT detections are those given by the source fitting
program.  Systematic errors are $<5\%$ for MOST but may be as high as
$30\%$ for GMRT due to the low elevation of the source.  Upper limits
on flux densities are three times the rms noise level measured in the
vicinity of the source position.

A radio measurement made with MOST on MJD 52443.625 gave a position for
\1543 of $15^{\rm h}47^{\rm m}8^{\rm s}.27$,
$-47\deg40\arcmin12\arcsec.8$ (J2000) with an error of $\sim3\arcsec$.
This is consistent with the positional measurements of the optical
counterpart from the UK Schmidt images on the Super COSMOS website in
the I--band: $15^{\rm h}47^{\rm m}8^{\rm s}.32$,
$-47\deg40\arcmin10\arcsec.8$ (J2000) with errors of $\sim0.25\arcsec$.

On MJD 52443, a 12--hour MOST survey observation detected a rapidly
rising radio flare.  During the first 6 hours (midpoint MJD 52443.375)
the mean flux density was $6 \pm 2$ mJy, while the second 6 hours (MJD
52443.625) gave a mean of $16 \pm 2$ mJy, suggesting that the source
had turned on quite suddenly.  This was consistent with a 30--minute
GMRT observation on the same day (MJD 52443.71) which gave a flux
density of $21.9 \pm 0.6$ mJy at 1026.75~MHz.  This radio flare
occurred two days before the peak of the X--ray outburst and decayed
quite rapidly.  However, the appearance of the radio flare at the low
frequencies of MOST and GMRT may be offset from the actual beginning
of the flare if the radio outburst was initially optically thick at
low frequencies.  Observations of the source on the following day gave
flux densities with upper limits of 2.7 mJy from a 12--hour
observation with MOST (MJD 52444.50), and 2.1 mJy from a 20--minute
observation with the GMRT (MJD 52444.71).  The GMRT also detected a
second radio flare with flux density 4.7 $\pm$ 0.3 mJy from a 2--hour
observation on MJD 52445.71 at 1026.75~MHz and 5.3 $\pm$ 0.4 mJy from
a 30--minute observation on MJD 52446.70 at 1286.75~MHz.  The source
went into a state of radio quiescence, with upper limits on the flux
density of 3.0 mJy from the MOST data taken on MJD 52459.37 and
52480.40, 0.6 mJy from the GMRT on MJD 52447.70 (1286.75~MHz), and 3.2
mJy on MJD 52480.62 (616.75~MHz).  Ten days after the X--ray
observations reported here were completed and after the source had
entered the {\it hard} state, the source was detected again at
$5.2 \pm 0.9$ mJy in a 12--hour MOST observation on MJD 52487.38, but
was quiescent on MJD 52496.33 with an upper limit of 2.4 mJy.

In a number of transient sources, the rising phase of an outburst is
spectrally hard in X--rays and accompanied by a radio flare (for
reviews, see Fender 2003, and McClintock \& Remillard 2003).  The
initial radio detections of \1543 are consistent with this picture.
In the most extreme cases, a hard X--ray flare may be associated with
blobs of material which are expelled at relativistic velocities, as
seen in radio images (e.g., Hannikainen et al.\ 2000).  The second
radio flare seen here (MJD 52445--6) suggests a somewhat more
complicated situation, with further, on--going activity.  Such
additional flaring has been seen in a number of other sources, though
on larger and longer scales; perhaps
the best--observed analogue is GRO J1655$-$40 (Hjellming and Rupen
1995), where the re--flares were associated with multiple relativistic
jet ejections, as well as substantial hard X--ray emission (Harmon et
al.\ 1995).  Unfortunately the radio observations discussed here do
not have sufficient angular resolution to reveal extended emission
from ejecta in \1543.

The radio detection of 4U 1543$-$47 on MJD 52487 may either indicate
another radio flare, or the transition to the {\it hard} state.
The ratio of the power--law flux to the total flux (see Fig.\ 2)
indicates a steep rise in the significance of the power--law spectral
component starting on or near MJD 52474.  The appearance of band
limited noise in the power--density spectra and QPOs on MJDs 54476 and
52477 also signal a transition to the {\it hard} state (see Fig.\
8).  The situation is not quite so simple however, because (fainter)
radio upper limits were derived on MJD 52480 and 52496, when the
source also appeared  to be in the same {\it hard} X--ray state.

\section{DISCUSSION} 

\subsection{Comparison with Earlier Outbursts of \1543}

It is problematic to compare the outbursts that occurred in 1971,
1983, 1992 and 2002 because of the differences in both instrumentation
and observing coverage (Tanaka \& Lewin 1995; Chen et al.\ 1997).
Nevertheless, we make a few comparisons, after setting aside the 1992
outburst, which was quite shortlived and observed only above 20~keV
(Harmon et al.\ 1992). For the three remaining outbursts, the spectra
were very soft when the source was bright, and the peak X--ray
intensities were 1.9 Crab (2--6~keV) in 1971, 4.0 Crab (3.7--7.5~keV)
in 1983, and 3.3 Crab (3.7--7.5~keV) on 20 June 2002 (Matilsky et al.\
1972; Kitamoto et al.\ 1984; this work).  During the bright outburst
in 1983, the luminosity above 1.5~keV was $2.5\pm0.3 \times
10^{39}$~erg s$^{-1}$ for D = 7.5~kpc (Kitamoto et al.\ 1984), which
nominally exceeds the Eddington luminosity of a 9.4~\msun~black hole
by a factor of 2.1. As indicated above, the peak luminosity during the
2002 outburst was within about 85\% of the overall maximum luminosity
reached in 1983.  However, the intensity in 2002 dropped rapidly with
a decay time (1/e) of $\approx13$ days (see Fig.\ 1).  By comparison,
the peak intensity decayed 4--5 times more slowly in 1971.

\subsection{Fe \Kalpha Emission Line Diagnostics}

Iron line features are likely produced when hard X--rays created in a
hot corona around a black hole irradiate a cold accretion disk and
generate a fluorescent Fe line (see, e.g., Shapiro \& Teukolsky 1983).
This emission line serves as a diagnostic for studies of gravitation
near the black hole.  The defining property of a black hole is its
event horizon, a surface through which particles and photons can fall
inward, but through which nothing can pass outward.  For a
non--rotating Schwarzschild black hole, the radius of the event
horizon is $R_S = 2(GM/c^2) \equiv 2~R_g$ and the innermost stable
circular orbit is $R_{ISCO} = 6~R_g$.  For a maximally--rotating Kerr
black hole, $R_K = R_{ISCO} \approx 1~R_g$.  Line emissions allow us
to probe the inner--disk accretion region and distinguish between
static and rotating black holes.

At its creation, a fluorescent Fe line is narrow and centered between
6.4~keV for neutral iron and 7~keV for hydrogenic iron.  However,
the observed line is severely broadened.  One broadening mechanism
that has been much discussed is Comptonization.  Model calculations
show that a line can be broadened due to Compton down--scattering in
an optically--thick cloud that is highly ionized (Czerny, Zbyszewska,
Raine 1991; see also Fabian et al.\ 1995).  In this scenario, an Fe
\Kalpha fluorescent line in an accretion disk could be broadened by
multiple Compton scatterings in an optically--thick corona.  This
model predicts a broad, {\it symmetric} line profile and a break in
the power--law continuum near 40~keV.  Numerous scatterings are
required to downshift an Fe \Kalpha photon by the 1~keV or more that
is frequently observed (e.g., Miller et al.\ 2002).  For example, even
in the extreme case of a $180^{\rm o}$ backscattering event, an Fe
\Kalpha photon loses only $\lesssim~0.2$~keV.

The Fe line can also be broadened due to relativistic distortions if
it is emitted from an accretion disk that is close to the black hole
(for reviews see Fabian et al.\ 2000; Reynolds \& Nowak 2003).  In the
models, relativistic beaming boosts the blue wing of the line while
attenuating the red wing, and gravitational redshifts displace the
profile down to lower energies.  The resultant profile is broad and
{\it asymmetric}, with its exact shape being a function of geometrical
parameters in the emitting regions, such as disk inner and outer
radii, and inclination angle.

The strongest emission lines we have observed demonstrate a very
asymmetric profile (a representative example is shown in Fig.\ 6a),
which argues against Comptonization as the broadening mechanism.
Furthermore, the small inclination angle of the source ($i = 21^{\rm
o}$) downplays the importance of Comptonization, in an effect
discussed by Petrucci et al.\ (2001).  These authors conclude that
decreasing the inclination angle has the same effect as decreasing the
optical depth, thereby decreasing the probability of a photon being
Comptonized.  The Laor model, which assumes a Kerr black hole and
accounts for strong Doppler shifts and gravitational redshifts,
naturally predicts an asymmetric profile that is similar to the one we
observe.  Indeed, the Laor model provides a good fit to all of our Fe
line data, including those intractable profiles for which the Gaussian
model fails (Fig.\ 5).  Thus, we conclude that the Fe line is
broadened largely by relativistic effects occurring in an accretion
disk.

Results from both the continuum and line fits suggest that the inner
edge of the accretion disk may extend within $6R_{g}$ (see Tables 1 \&
2, and Figures 3b and 4), which may serve as an indication for a non--zero spin
parameter.  To better assess the possibility of black hole spin in
\1543, we examined the broad Fe line profile observed in 1983 with the
superior resolution of the {\it EXOSAT}/GSPC (see \S1).  We chose to
fit this spectrum with the same models herein applied to our {\it
RXTE} observations, rather than the Comptonization model fitted
previously by the authors.  As with many of our spectra, the Gaussian
model fails to fit the data, and a Laor line model is required
(although a smeared edge is not).  To enable a direct comparison to
both our profiles (e.g., Fig.\ 6) and the profile featured by van der
Woerd et al.\ (1989), we show in Figure 10 both the fit residuals and
the ratio data/model.  The residuals are remarkably similar to those
obtained by van der Woerd et al.\ using the Comptonization model,
which indicates that the line profile is largely {\it independent} of
the particular continuum model.  Moreover, the line profile appears to be
double--peaked, consistent with the best--studied lines in AGN.
Fitting the {\it EXOSAT} spectrum with the Laor model suggests a line
of only moderate strength ($EW = 180\pm 60$~eV), but from a high
charge state (E $=6.87-7$~keV) compared to the {\it RXTE} results
summarized in Table~2.  Significantly, the {\it EXOSAT} Fe \Kalpha
line more strongly requires emission from within $6~R_{g}$: $\rho =
3.3\pm 0.6$ in units of $R_{g}$.

Van der Woerd et al.\ (1989) fit the Fe line with a Gaussian and found
the central energy of the line to be $5.93 \pm 0.24$~keV.  We also
find that the line center shifts downward to a mean value of $\approx
5.4$~keV when we fit the {\it RXTE} line profiles allowing the
Gaussian to float down to a lower energy limit of 5~keV.   These fits
provided reduced chi--squared values similar to the ones obtained by
fitting a Laor model.  However, as we noted earlier, because the Fe
lines we observed are obviously asymmetric (e.g., Fig.\ 6a), we favor
the more physical Laor model to an unbounded Gaussian model.  This
choice is further supported by the Fe line profiles observed for other
sources at higher resolution, which show that the lines are generally
asymmetric, as well as by theoretical considerations which favor
relativistic asymmetric broadening.  

It should be noted that this analysis marks the first published
attempt to systematically fit numerous Galactic black hole X--ray
spectra with a relativistic emission line model, while covering a
range in luminosity with variations by a factor $> 30$ during a single
outburst.  The brightness and low inclination of \1543 presents a
special opportunity to study the source in greater detail.  Laor
models clearly have the 
potential to provide important constraints on the inner accretion
flow and the nature of the black hole itself.  Prior efforts to fit
numerous spectra spanning transient black hole outbursts have
primarily relied on Gaussian models, and may have missed valuable
information.

\subsection{Constraining the Accretion Flow Geometry}

The disk blackbody model parameters summarized in Table~1 indicate
that the inner disk edge likely remained relatively stable within the
marginally stable circular orbit of a black hole despite a decline in
total flux by a factor of 32 and a drop in the color temperature of
the inner disk by a factor of 2.  This constancy of $r_{col}$ over a
wide range of mass accretion rates has been observed for a number of
X--ray novae during their decline (e.g., Tanaka \& Lewin 1995; Sobczak
et al.\ 2000).  When the power--law flux is low or moderate, we
regard the values of $r_{col}$ in Table~1 as a good indicator of
relative changes in the inner disk, as well as an approximate
indicator of the inner disk radius.  However, these values do not
include relativistic corrections (e.g., Zhang, Cui \& Chen 1997) or
corrections for the effects of electron scattering (\S3), which are
themselves quite uncertain (Merloni et al.\ 2000).  Also, the errors
in $r_{col}$ in Table 1, have not been calculated to include errors in
the distance, the mass of the black hole, and the binary inclination.
The former two uncertainties (see \S1) are dominant and translate into
an error of 25\% on $r_{col}$.  Not only was $r_{col}$ fairly stable
during all of our observations, the values of $\rho$ inferred from the
Laor model were similarly stable throughout (Table~2, Fig.\ 4).

The fact that the power--law and Fe~K$\alpha$ line fluxes are
correlated on long time scales (Fig.\ 2cd) confirms the basic disk
reflection picture described by George \& Fabian (1991) and the models
developed by Zdziarski et al.\ (2003).  The line flux peaks at the
start of the outburst, during the same observations when the disk
color temperature is highest.  However, during this early phase the
power--law flux is not at its absolute peak value; moreover, it reaches a
much higher fraction of the total flux at later phases of the outburst
(Fig.\ 2e).  Assuming that the hard X--ray emission arises in an
optically thin corona, the early maximum in the Fe~K$\alpha$ line
(despite the modest level of the power--law flux) may indicate that
the corona was more centrally--concentrated and/or it illuminated the
disk more centrally early in the outburst.  A similar suggestion was
made to explain differences in the Fe~K$\alpha$ emission line profile
in the Seyfert--1 galaxy MCG--6-30-15 (Iwasawa et al.\ 1996).

As noted previously, the binary inclination of \1543 is very
low, $i=21^{\rm o}$.  If we assume that the inner disk is seen
at a similar inclination (a reasonable assumption given the success of
fits with the Laor model which assume this inclination), then the axis
of any jet outflow may be close to our line of sight.  Recently, it
has been suggested that the hard X--ray component in Galactic black
holes may due to direct synchrotron emission and/or synchrotron
self--Comptonization in the jet (Markoff, Falcke, \& Fender 2001;
Markoff et al.\ 2003).  It is therefore particularly interesting to
assess the role of a jet in \1543.  

Some aspects of our results argue against associating the hard X--ray
emission in \1543 with a jet, at least in this phase of the
outburst.  First, the power--law index is rather soft; synchrotron
radiation from a jet with a flat to ``inverted'' radio spectrum should
be spectrally harder.  Also, the Lorentz factors required to
generate X--ray synchrotron emission in a jet would beam the hard
X--ray emission in a narrow cone along the jet axis and away from the
disk.  This is inconsistent with our observation of a broad and
intense Fe K$\alpha$ emission line which is very likely produced by
hard X--ray emission shining ``down'' onto the accretion disk.  For
example, the X--ray reflection models of George \& Fabian (1991)
predict an Fe K$\alpha$ equivalent width of $\lesssim120$~eV for $i =
21^{\rm o}$ and the steep power--law indices observed by us
($\Gamma~>~2.3$; see their Fig.\ 14).  Averaged over all of our
observations, the equivalent width of the Fe \Kalpha line is
$\approx2$ times this predicted value, and it is occasionally 3--4
times greater (Table~2).  Clearly, the large line equivalent widths we
measure demand that a high fraction of the total hard flux be
intercepted by the disk, which is incompatible with the emission being
beamed away from the disk.  

\section{SUMMARY AND CONCLUSIONS}

We have analyzed the spectral and timing properties \1543
during the bright phase of its 2002 outburst using observations
obtained with the {\it RXTE}/PCA.  The outburst, which exhibited a
fast rise and an exponential decay, was dominated throughout our
observations by soft thermal emission from the accretion disk; the
source was generally in the {\it thermal dominant} state.  However,
midway through the decay phase, the source entered the {\it steep
power--law} state and strong low--frequency QPOs were detected with
obvious harmonic features.  As frequently observed for this state, the
power--law flux increased to a maximum and the radius of the inner disk
inferred from the color temperature, $r_{col}$, decreased.  Near the
end of our observations, a low--frequency QPO reappeared and the
power--law spectrum hardened as the source entered the {\it hard}
state.  

A radio flare was detected near the peak of the X--ray
outburst by MOST and GMRT with a peak flux density of $21.9 \pm 0.6$
mJy.  Smaller flares were also detected just after the X--ray peak, as
well as after the source entered the {\it hard} state.  The radio
observations did not have sufficient angular resolution to reveal
evidence of extended emission from ejected blobs.

The most striking feature of the X--ray spectrum is a broad,
asymmetric Fe~K$\alpha$ emission line.  We fit this line in each of
our spectra using the Laor relativistic line model, which generally
provided a much better fit to our data than Gaussian models.  Despite
the low resolution of the {\it RXTE}/PCA, about half the spectral fits
suggest that \1543 contains a black hole with non--zero angular
momentum.  The higher resolution {\it EXOSAT} spectrum (Van der Woerd
et al.\ 1989) lends further, significant support to the suggestion
that \1543 contains a Kerr black hole.

Although numerous transient black hole outbursts have been observed
extensively with the {\it RXTE}/PCA and prior observatories such as
{\it Ginga}, this is the first systematic effort to fit a relativistic
line model to numerous spectra which span an order of magnitude in
flux.  Our results suggest that proportional counter data obtained in
earlier synoptic studies of black hole binaries can be reanalyzed
using a relativistic emission line model to obtain important
constraints on the nature of the black hole and the geometry of the
accretion flow.

\acknowledgments This work is supported in part by NASA Grant
NAG5--10813. This work has made use of the information and tools
available at the HEASARC website, operated by GSFC for NASA.  We
thank the  staff of GMRT that made these observations possible. GMRT
is run by the  National Centre of Radio Astrophysics of the Tata
Institute of Fundamental  Research.  MOST is operated by the
University of Sydney and supported in part by grants from the
Australian Research Council.

\clearpage

\begin{deluxetable}{cccccccccccc}
\tabletypesize{\tiny}
\tablewidth{565pt}
\tablecaption{\textbf{PCA Data and Broadband Spectral Parameters for
  \1543}}
\tablehead{
 \colhead{Obs} &    	   \colhead{Date} &   	      \colhead{MJD\tablenotemark{a}} &
 \colhead{Obs.} &          \colhead{$T_{col}$\tablenotemark{b}}&  \colhead{$r_{col}$\tablenotemark{c}}&
 \colhead{Photon} &	   \colhead{Power--Law\tablenotemark{d}}&  \colhead{Fe Edge\tablenotemark{e}} &
 \colhead{$\tau_{Fe}$\tablenotemark{e}}&\colhead{Flux \X$10^{-10}$\tablenotemark{f}}&     \colhead{\redchi2}\\
 \colhead{No.} &    	   \colhead{(mmdd)} & 	      \colhead{} &
 \colhead{Time (s)} &      \colhead{(keV)} &  	      \colhead{($R_g$)} &
 \colhead{Index} &         \colhead{Norm} &           \colhead{(keV)} &
 \colhead{}    &           \colhead{(\ergcm2s)} &     \colhead{(40 dof)}}
\startdata
1   & 0617 & 52442.83 &2544 & 0.90$^{+0.01}_{-0.02}$ & 4.50$^{+0.20}_{-0.21}$ & 2.55$^{+0.11}_{-0.09}$ &  5.11$^{+ 2.14}_{- 1.31}$ &7.48$^{+0.37}_{-0.38}$ & 1.08$^{+0.54}_{-0.48}$ & 242.74$^{+ 0.60}_{-1.61}$ &  0.97\\
2   & 0617 & 52442.90 &5216 & 0.91$^{+0.02}_{-0.01}$ & 4.45$^{+0.18}_{-0.22}$ & 2.52$^{+0.10}_{-0.07}$ &  5.06$^{+ 1.85}_{- 1.10}$ &7.55$^{+0.33}_{-0.39}$ & 1.10$^{+0.51}_{-0.41}$ & 257.69$^{+ 0.61}_{-1.17}$ &  0.87\\
3   & 0618 & 52443.26 &2992 & 0.94$^{+0.01}_{-0.01}$ & 4.68$^{+0.17}_{-0.17}$ & 2.65$^{+0.09}_{-0.08}$ &  5.54$^{+ 1.83}_{- 1.18}$ &7.36$^{+0.31}_{-0.26}$ & 1.37$^{+0.37}_{-0.37}$ & 314.48$^{+ 1.11}_{-1.28}$ &  0.91\\
4   & 0619 & 52444.45 &3552 & 1.02$^{+0.01}_{-0.01}$ & 4.49$^{+0.09}_{-0.13}$ & 3.71$^{+0.21}_{-0.11}$ & 33.69$^{+27.47}_{- 9.07}$ &7.63$^{+0.19}_{-0.16}$ & 1.63$^{+0.31}_{-0.22}$ & 453.91$^{+ 1.01}_{-2.07}$ &  1.63\\
5   & 0619 & 52444.52 &3744 & 1.03$^{+0.01}_{-0.01}$ & 4.39$^{+0.11}_{-0.13}$ & 3.85$^{+0.16}_{-0.12}$ & 52.75$^{+29.98}_{-14.72}$ &7.59$^{+0.18}_{-0.07}$ & 1.73$^{+0.26}_{-0.24}$ & 459.97$^{+ 1.46}_{-2.10}$ &  1.48\\
6   & 0620 & 52445.57 &3968 & 1.02$^{+0.01}_{-0.01}$ & 4.61$^{+0.15}_{-0.14}$ & 2.93$^{+0.09}_{-0.07}$ & 16.52$^{+ 5.44}_{- 3.25}$ &7.76$^{+0.31}_{-0.26}$ & 1.20$^{+0.32}_{-0.29}$ & 501.46$^{+ 1.29}_{-2.64}$ &  1.30\\
7   & 0620 & 52445.64 &4144 & 1.04$^{+0.01}_{-0.01}$ & 4.49$^{+0.12}_{-0.14}$ & 2.73$^{+0.08}_{-0.06}$ &  9.53$^{+ 3.58}_{- 0.80}$ &7.85$^{+0.38}_{-0.30}$ & 0.96$^{+0.35}_{-0.25}$ & 523.06$^{+ 1.51}_{-2.36}$ &  1.35\\
8   & 0621 & 52446.07 &4432 & 1.04$^{+0.01}_{-0.01}$ & 4.70$^{+0.13}_{-0.12}$ & 3.41$^{+0.16}_{-0.11}$ & 19.26$^{+ 5.86}_{- 4.96}$ &7.65$^{+0.18}_{-0.20}$ & 1.41$^{+0.28}_{-0.19}$ & 491.25$^{+ 1.24}_{-2.97}$ &  1.41\\
9   & 0622 & 52447.00 &4560 & 1.03$^{+0.01}_{-0.01}$ & 4.72$^{+0.17}_{-0.16}$ & 3.15$^{+0.15}_{-0.04}$ & 11.24$^{+ 6.09}_{- 2.33}$ &7.56$^{+0.24}_{-0.20}$ & 1.37$^{+0.34}_{-0.26}$ & 470.82$^{+ 1.52}_{-1.96}$ &  1.24\\
10  & 0623 & 52448.06 &3312 & 1.02$^{+0.01}_{-0.01}$ & 4.45$^{+0.14}_{-0.18}$ & 2.91$^{+0.13}_{-0.09}$ & 6.446$^{+ 2.91}_{- 1.47}$ &7.59$^{+0.27}_{-0.22}$ & 1.35$^{+0.43}_{-0.28}$ & 438.86$^{+ 1.34}_{-1.49}$ &  1.42\\
11  & 0624 & 52449.11 &2064 & 1.01$^{+0.01}_{-0.01}$ & 4.51$^{+0.20}_{-0.16}$ & 2.60$^{+0.10}_{-0.09}$ & 5.142$^{+ 1.88}_{- 1.25}$ &7.69$^{+0.32}_{-0.29}$ & 1.19$^{+0.33}_{-0.32}$ & 420.92$^{+ 1.58}_{-1.60}$ &  1.14\\
12  & 0625 & 52450.25 &1680 & 0.97$^{+0.01}_{-0.01}$ & 4.52$^{+0.19}_{-0.20}$ & 2.68$^{+0.10}_{-0.08}$ &  6.18$^{+ 1.93}_{- 1.39}$ &7.44$^{+0.28}_{-0.14}$ & 1.29$^{+0.32}_{-0.30}$ & 375.10$^{+ 1.56}_{-1.81}$ &  0.75\\
13  & 0626 & 52451.57 &1856 & 0.95$^{+0.01}_{-0.01}$ & 4.50$^{+0.22}_{-0.19}$ & 2.75$^{+0.10}_{-0.09}$ &  6.76$^{+ 2.22}_{- 1.69}$ &7.29$^{+0.26}_{-0.19}$ & 1.63$^{+0.35}_{-0.41}$ & 333.55$^{+ 0.98}_{-1.54}$ &  0.79\\
14  & 0627 & 52452.69 &2496 & 0.97$^{+0.01}_{-0.01}$ & 4.63$^{+0.19}_{-0.17}$ & 2.48$^{+0.05}_{-0.08}$ &  3.34$^{+ 1.04}_{- 0.69}$ &7.38$^{+0.29}_{-0.26}$ & 1.44$^{+0.32}_{-0.34}$ & 329.47$^{+ 1.18}_{-1.41}$ &  0.69\\
15  & 0628 & 52453.50 &1344 & 0.93$^{+0.01}_{-0.02}$ & 4.60$^{+0.20}_{-0.17}$ & 2.55$^{+0.10}_{-0.08}$ &  7.99$^{+ 2.93}_{- 1.85}$ &7.99$^{+0.39}_{-0.38}$ & 0.82$^{+0.47}_{-0.39}$ & 314.14$^{+ 1.06}_{-1.43}$ &  0.71\\
16  & 0628 & 52453.55 &3584 & 0.93$^{+0.01}_{-0.02}$ & 3.90$^{+0.20}_{-0.17}$ & 2.59$^{+0.10}_{-0.08}$ &  7.57$^{+ 2.96}_{- 1.67}$ &7.64$^{+0.37}_{-0.40}$ & 0.95$^{+0.52}_{-0.40}$ & 306.95$^{+ 0.99}_{-1.26}$ &  0.60\\
17  & 0629 & 52454.60 &2048 & 0.91$^{+0.01}_{-0.02}$ & 3.88$^{+0.19}_{-0.19}$ & 2.57$^{+0.12}_{-0.08}$ &  7.06$^{+ 3.46}_{- 1.56}$ &7.65$^{+0.33}_{-0.35}$ & 1.07$^{+0.63}_{-0.40}$ & 280.12$^{+ 0.86}_{-1.62}$ &  0.63\\
18  & 0630 & 52455.13 &656  & 0.91$^{+0.01}_{-0.02}$ & 3.34$^{+0.22}_{-0.25}$ & 2.41$^{+0.21}_{-0.14}$ &  2.73$^{+ 2.58}_{- 0.96}$ &7.41$^{+1.89}_{-0.31}$ & 0.52$^{+0.95}_{-0.52}$ & 253.72$^{+ 0.88}_{-1.30}$ &  0.95\\
19  & 0701 & 52456.77 &4944 & 0.88$^{+0.01}_{-0.02}$ & 3.41$^{+0.27}_{-0.19}$ & 2.46$^{+0.12}_{-0.09}$ &  3.08$^{+ 1.49}_{- 0.76}$ &7.57$^{+0.43}_{-0.39}$ & 0.90$^{+0.63}_{-0.47}$ & 215.46$^{+ 0.95}_{-1.36}$ &  0.57\\
20  & 0702 & 52457.83 &1776 & 0.90$^{+0.02}_{-0.02}$ & 3.27$^{+0.25}_{-0.21}$ & 2.74$^{+0.12}_{-0.07}$ & 10.62$^{+ 5.04}_{- 2.14}$ &7.85$^{+0.29}_{-0.36}$ & 1.15$^{+0.62}_{-0.36}$ & 222.76$^{+ 0.65}_{-1.10}$ &  0.97\\
21  & 0703 & 52458.44 &1904 & 0.89$^{+0.02}_{-0.02}$ & 3.72$^{+0.22}_{-0.19}$ & 2.77$^{+0.11}_{-0.09}$ & 11.55$^{+ 4.68}_{- 2.78}$ &7.70$^{+0.30}_{-0.43}$ & 1.32$^{+0.57}_{-0.46}$ & 210.91$^{+ 0.64}_{-0.81}$ &  0.85\\
22  & 0704 & 52459.08 &1152 & 0.90$^{+0.02}_{-0.03}$ & 3.78$^{+0.22}_{-0.18}$ & 2.72$^{+0.11}_{-0.07}$ & 16.05$^{+ 6.39}_{- 3.39}$ &7.67$^{+0.31}_{-0.37}$ & 1.30$^{+0.62}_{-0.41}$ & 227.65$^{+ 0.47}_{-1.04}$ &  0.69\\
23  & 0705 & 52460.49 &1216 & 0.87$^{+0.02}_{-0.03}$ & 4.52$^{+0.29}_{-0.16}$ & 2.51$^{+0.10}_{-0.05}$ &  9.03$^{+ 3.32}_{- 1.45}$ &7.94$^{+0.32}_{-0.41}$ & 1.13$^{+0.59}_{-0.30}$ & 196.09$^{+ 0.53}_{-0.85}$ &  0.66\\
24  & 0706 & 52461.41 &880  & 0.86$^{+0.03}_{-0.03}$ & 4.80$^{+0.23}_{-0.20}$ & 2.64$^{+0.10}_{-0.08}$ & 11.61$^{+ 4.48}_{- 2.70}$ &7.70$^{+0.39}_{-0.35}$ & 1.40$^{+0.59}_{-0.47}$ & 180.08$^{+ 0.37}_{-1.14}$ &  0.92\\
25  & 0707 & 52462.57 &4928 & 0.83$^{+0.02}_{-0.02}$ & 4.58$^{+0.22}_{-0.19}$ & 2.59$^{+0.07}_{-0.06}$ &  5.47$^{+ 1.46}_{- 1.01}$ &7.89$^{+0.25}_{-0.24}$ & 1.53$^{+0.43}_{-0.36}$ & 130.59$^{+ 0.32}_{-0.54}$ &  0.94\\
26  & 0708 & 52463.67 &4816 & 0.81$^{+0.02}_{-0.02}$ & 4.68$^{+0.36}_{-0.30}$ & 2.61$^{+0.09}_{-0.06}$ &  4.05$^{+ 1.26}_{- 0.73}$ &7.99$^{+0.24}_{-0.27}$ & 1.55$^{+0.47}_{-0.34}$ & 107.84$^{+ 0.39}_{-0.46}$ &  0.71\\
27  & 0709 & 52464.45 &2768 & 0.76$^{+0.01}_{-0.02}$ & 4.58$^{+0.23}_{-0.18}$ & 2.32$^{+0.24}_{-0.11}$ &  0.61$^{+ 0.68}_{- 0.17}$ &7.95$^{+0.51}_{-0.56}$ & 1.23$^{+1.22}_{-0.51}$ & 75.087$^{+ 0.17}_{-0.45}$ &  0.73\\
28  & 0710 & 52465.08 &1200 & 0.73$^{+0.01}_{-0.01}$ & 4.84$^{+0.30}_{-0.22}$ & 2.40$^{+0.24}_{-0.15}$ &  0.75$^{+ 0.82}_{- 0.26}$ &7.94$^{+0.61}_{-0.59}$ & 1.56$^{+1.19}_{-0.67}$ &  69.27$^{+ 0.24}_{-0.51}$ &  0.82\\
29  & 0711 & 52466.34 &2720 & 0.72$^{+0.01}_{-0.01}$ & 4.77$^{+0.25}_{-0.18}$ & 2.35$^{+0.10}_{-0.09}$ &  0.79$^{+ 0.28}_{- 0.17}$ &8.08$^{+0.31}_{-0.32}$ & 1.82$^{+0.47}_{-0.39}$ &  59.62$^{+ 0.20}_{-0.37}$ &  0.99\\
30  & 0712 & 52467.20 &1888 & 0.70$^{+0.02}_{-0.02}$ & 5.04$^{+0.32}_{-0.23}$ & 2.62$^{+0.20}_{-0.25}$ &  1.35$^{+ 1.10}_{- 0.72}$ &7.43$^{+1.13}_{-0.33}$ & 2.06$^{+1.12}_{-1.20}$ &  51.91$^{+ 0.26}_{-0.35}$ &  0.87\\
31  & 0713 & 52468.66 &3328 & 0.69$^{+0.01}_{-0.01}$ & 5.00$^{+0.42}_{-0.24}$ & 2.31$^{+0.12}_{-0.14}$ &  0.34$^{+ 0.35}_{- 0.11}$ &8.01$^{+0.45}_{-0.55}$ & 2.08$^{+1.12}_{-0.57}$ &  41.03$^{+ 0.14}_{-0.19}$ &  0.61\\
32  & 0713 & 52468.73 &1648 & 0.68$^{+0.01}_{-0.01}$ & 4.99$^{+0.19}_{-0.26}$ & 2.58$^{+0.23}_{-0.20}$ &  0.68$^{+ 0.56}_{- 0.30}$ &7.63$^{+0.42}_{-0.43}$ & 3.13$^{+0.84}_{-0.92}$ &  39.84$^{+ 0.17}_{-0.26}$ &  0.87\\
33  & 0714 & 52469.24 &1760 & 0.67$^{+0.01}_{-0.01}$ & 4.76$^{+0.29}_{-0.33}$ & 2.47$^{+0.25}_{-0.16}$ &  0.47$^{+ 0.45}_{- 0.15}$ &8.04$^{+0.48}_{-0.63}$ & 1.85$^{+0.94}_{-0.57}$ &  37.64$^{+ 0.20}_{-0.26}$ &  0.71\\
34  & 0714 & 52469.31 &3104 & 0.65$^{+0.01}_{-0.01}$ & 5.22$^{+0.34}_{-0.30}$ & 2.47$^{+0.20}_{-0.14}$ &  0.57$^{+ 0.44}_{- 0.19}$ &7.66$^{+0.35}_{-0.40}$ & 2.78$^{+0.95}_{-0.66}$ &  35.64$^{+ 0.17}_{-0.20}$ &  0.82\\
35  & 0715 & 52470.24 &1376 & 0.65$^{+0.01}_{-0.02}$ & 5.26$^{+0.27}_{-0.30}$ & 2.52$^{+0.33}_{-0.01}$ &  0.55$^{+ 0.76}_{- 0.24}$ &7.49$^{+0.51}_{-0.39}$ & 3.21$^{+1.45}_{-1.04}$ &  31.78$^{+ 0.13}_{-0.22}$ &  0.89\\
36  & 0715 & 52470.30 &3472 & 0.64$^{+0.01}_{-0.01}$ & 5.13$^{+0.35}_{-0.36}$ & 2.66$^{+0.12}_{-0.15}$ &  0.85$^{+ 0.33}_{- 0.32}$ &7.27$^{+0.27}_{-0.17}$ & 3.75$^{+0.47}_{-0.92}$ &  31.87$^{+ 0.16}_{-0.21}$ &  0.65\\
37  & 0715 & 52470.98 &592  & 0.62$^{+0.01}_{-0.01}$ & 5.19$^{+0.24}_{-0.31}$ & 3.84$^{+0.19}_{-0.22}$ &  5.57$^{+ 3.61}_{- 3.40}$ &7.18$^{+0.81}_{-0.08}$ & 1.69$^{+1.25}_{-1.45}$ &  24.09$^{+ 0.02}_{-0.68}$ &  0.99\\
38  & 0716 & 52471.02 &1072 & 0.62$^{+0.01}_{-0.01}$ & 5.19$^{+0.31}_{-0.34}$ & 3.09$^{+0.28}_{-0.27}$ &  1.44$^{+ 1.46}_{- 0.80}$ &7.17$^{+0.41}_{-0.07}$ & 4.04$^{+1.33}_{-1.74}$ &  26.56$^{+ 0.07}_{-0.31}$ &  0.93\\
39  & 0717 & 52472.62 &2048 & 0.59$^{+0.01}_{-0.02}$ & 4.77$^{+0.49}_{-0.51}$ & 2.75$^{+0.21}_{-0.18}$ &  0.76$^{+ 0.32}_{- 0.29}$ &7.25$^{+0.46}_{-0.15}$ & 2.89$^{+0.65}_{-0.85}$ &  20.58$^{+ 0.12}_{-0.26}$ &  0.63\\
40  & 0717 & 52472.69 &1712 & 0.60$^{+0.01}_{-0.01}$ & 4.25$^{+0.29}_{-0.20}$ & 2.54$^{+0.23}_{-0.21}$ &  0.53$^{+ 0.42}_{- 0.23}$ &7.45$^{+0.48}_{-0.35}$ & 3.38$^{+0.87}_{-0.94}$ &  20.88$^{+ 0.07}_{-0.18}$ &  1.11\\
41  & 0718 & 52473.20 &1760 & 0.59$^{+0.01}_{-0.00}$ & 4.65$^{+0.52}_{-0.98}$ & 2.67$^{+0.08}_{-0.10}$ &  0.93$^{+ 0.19}_{- 0.23}$ &7.10$^{+0.27}_{-0.00}$ & 2.71$^{+0.30}_{-0.79}$ &  21.02$^{+ 0.08}_{-0.14}$ &  0.84\\
42  & 0718 & 52473.27 &2864 & 0.59$^{+0.01}_{-0.01}$ & 4.39$^{+0.68}_{-0.64}$ & 2.60$^{+0.12}_{-0.13}$ &  0.72$^{+ 0.25}_{- 0.22}$ &7.26$^{+0.33}_{-0.16}$ & 3.09$^{+0.49}_{-0.70}$ &  20.47$^{+ 0.08}_{-0.16}$ &  0.87\\
43  & 0719 & 52474.18 &1104 & 0.58$^{+0.02}_{-0.02}$ & 4.45$^{+0.10}_{-0.13}$ & 2.56$^{+0.13}_{-0.14}$ &  1.02$^{+ 0.39}_{- 0.32}$ &7.32$^{+0.44}_{-0.22}$ & 2.75$^{+0.49}_{-0.66}$ &  20.35$^{+ 0.12}_{-0.28}$ &  0.86\\
44  & 0719 & 52474.94 &880  & 0.57$^{+0.03}_{-0.01}$ & 4.49$^{+0.11}_{-0.12}$ & 2.77$^{+0.04}_{-0.12}$ &  1.76$^{+ 0.20}_{- 0.47}$ &7.10$^{+0.24}_{-0.00}$ & 1.72$^{+0.26}_{-0.22}$ &  19.32$^{+ 0.05}_{-0.24}$ &  1.19\\
45  & 0721 & 52476.02 &1312 & 0.53$^{+0.04}_{-0.02}$ & 4.53$^{+0.12}_{-0.11}$ & 2.48$^{+0.02}_{-0.08}$ & 1.408$^{+ 0.12}_{- 0.30}$ &7.10$^{+0.33}_{-0.00}$ & 2.41$^{+0.19}_{-0.25}$ &  20.75$^{+ 0.04}_{-0.28}$ &  1.17\\
46  & 0721 & 52476.08 &1760 & 0.54$^{+0.03}_{-0.03}$ & 4.48$^{+0.13}_{-0.13}$ & 2.45$^{+0.04}_{-0.07}$ & 1.259$^{+ 0.16}_{- 0.22}$ &7.10$^{+0.27}_{-0.00}$ & 2.31$^{+0.32}_{-0.42}$ &  20.12$^{+ 0.06}_{-0.17}$ &  0.93\\
47  & 0721 & 52476.15 &3520 & 0.52$^{+0.03}_{-0.01}$ & 4.55$^{+0.52}_{-0.92}$ & 2.44$^{+0.02}_{-0.06}$ & 1.256$^{+ 0.07}_{- 0.20}$ &7.10$^{+0.18}_{-0.00}$ & 2.88$^{+0.08}_{-0.38}$ &  19.78$^{+ 0.07}_{-0.15}$ &  0.78\\
48  & 0722 & 52477.23 &1408 & 0.51$^{+0.04}_{-0.01}$ & 4.15$^{+0.73}_{-0.62}$ & 2.41$^{+0.04}_{-0.06}$ &  1.03$^{+ 0.09}_{- 0.20}$ &7.10$^{+0.24}_{-0.00}$ & 2.80$^{+0.15}_{-0.50}$ &  16.99$^{+ 0.05}_{-0.31}$ &  0.84\\
49  & 0722 & 52477.30 &3536 & 0.52$^{+0.03}_{-0.02}$ & 4.64$^{+0.44}_{-0.76}$ & 2.40$^{+0.04}_{-0.06}$ &  0.94$^{+ 0.11}_{- 0.14}$ &7.15$^{+0.21}_{-0.05}$ & 2.74$^{+0.26}_{-0.35}$ &  16.35$^{+ 0.05}_{-0.10}$ &  0.86\\
\enddata
\tablecomments{Output of spectral parameters from an input model
  consisting of a multi--color blackbody, power--law, Laor line
  emission, and a smeared absorption Fe edge.  $N_H = 4 \times
  10^{21}$ cm$^{-2}$ was assumed throughout.}
\tablenotetext{a}{Midpoint of observation.}
\tablenotetext{b}{Temperature at the inner disk radius.}
\tablenotetext{c}{$r_{col}$ = inner radius in units of $R_g$,
  where $R_g=GM/c^2$, and M = 9.4~\msun, and we have also assumed
  $i=21\deg$ and $d=7.5$~kpc.}  
\tablenotetext{d}{Photons keV$^{-1}$ cm$^{-2}$ s$^{-1}$ at 1~keV.}
\tablenotetext{e}{Fixed width at 7~keV using the `smedge' model
  in XSPEC.} 
\tablenotetext{f}{Unabsorbed flux in the 2.9--25~keV
  band.  The unabsorbed fluxes in the 0.5--10~keV band are 3.4--15.0
  time higher than fluxes in the 3--25~keV band.}
\end{deluxetable}

\begin{deluxetable}{cccccccc}
\tabletypesize{\tiny}
\tablewidth{7truein}
\tablecaption{\textbf{Fe Line Spectral Parameters for \1543 -- Laor Model\label{data.table}} }
\tablehead{
 \colhead{Obs} &    	  \colhead{Date} &           \colhead{MJD\tablenotemark{a}} &   
 \colhead{Energy\tablenotemark{b}} &   \colhead{$\rho$\tablenotemark{c}}&      \colhead{Normalization\tablenotemark{d}} &
 \colhead{Equivalent} &	  \colhead{Flux \X$10^{-10}$\tablenotemark{e}} \\ 
 \colhead{No.} &    	  \colhead{(mmdd)} & 	     \colhead{} &           
 \colhead{(keV)} &        \colhead{($R_g$)} &  	     \colhead{} &
 \colhead{Width (eV)} &        \colhead{(\ergcm2s)}
}
\startdata        
1  & 0617 & 52442.83 & 6.58$^{+0.39}_{-0.18}$ & 3.54$^{+396.46}_{-2.30}$ & 0.019$^{+0.014}_{-0.019}$ & 129$^{+ 99}_{-129}$ & 1.55$^{+0.24}_{-0.26}$ \\ 
2  & 0617 & 52442.90 & 6.68$^{+0.30}_{-0.28}$ & 3.37$^{+396.63}_{-2.14}$ & 0.020$^{+0.014}_{-0.020}$ & 136$^{+ 97}_{-136}$ & 1.67$^{+0.27}_{-0.28}$ \\ 
3  & 0618 & 52443.26 & 6.44$^{+0.53}_{-0.04}$ & 3.70$^{+396.30}_{-2.47}$ & 0.024$^{+0.019}_{-0.024}$ & 110$^{+ 88}_{-110}$ & 1.98$^{+0.44}_{-0.47}$ \\ 
4  & 0619 & 52444.45 & 6.41$^{+0.18}_{-0.01}$ & 3.70$^{+  2.37}_{-2.47}$ & 0.066$^{+0.037}_{-0.023}$ & 181$^{+103}_{- 64}$ & 5.29$^{+0.78}_{-0.72}$ \\ 
5  & 0619 & 52444.52 & 6.40$^{+0.16}_{-0.00}$ & 3.54$^{+  2.00}_{-2.30}$ & 0.071$^{+0.013}_{-0.025}$ & 188$^{+ 35}_{- 67}$ & 5.69$^{+0.57}_{-0.70}$ \\ 
6  & 0620 & 52445.57 & 6.56$^{+0.22}_{-0.16}$ & 4.03$^{+  5.41}_{-2.80}$ & 0.068$^{+0.026}_{-0.028}$ & 186$^{+ 72}_{- 77}$ & 5.60$^{+0.77}_{-0.58}$ \\ 
7  & 0620 & 52445.64 & 6.75$^{+0.17}_{-0.25}$ & 3.70$^{+  5.79}_{-2.47}$ & 0.074$^{+0.027}_{-0.028}$ & 222$^{+ 81}_{- 83}$ & 6.32$^{+0.73}_{-0.79}$ \\ 
8  & 0621 & 52446.07 & 6.45$^{+0.26}_{-0.06}$ & 4.03$^{+  2.63}_{-2.80}$ & 0.067$^{+0.009}_{-0.040}$ & 182$^{+ 26}_{-108}$ & 5.52$^{+0.68}_{-0.57}$ \\
9  & 0622 & 52447.00 & 6.42$^{+0.28}_{-0.02}$ & 4.03$^{+  5.26}_{-2.80}$ & 0.060$^{+0.026}_{-0.029}$ & 165$^{+ 70}_{- 80}$ & 4.93$^{+0.55}_{-0.69}$ \\
10 & 0623 & 52448.06 & 6.49$^{+0.23}_{-0.09}$ & 4.03$^{+  5.43}_{-2.80}$ & 0.056$^{+0.023}_{-0.013}$ & 170$^{+ 71}_{- 41}$ & 4.55$^{+0.56}_{-0.66}$ \\
11 & 0624 & 52449.11 & 6.57$^{+0.24}_{-0.18}$ & 3.70$^{+  8.55}_{-2.47}$ & 0.049$^{+0.021}_{-0.025}$ & 166$^{+ 73}_{- 85}$ & 4.04$^{+0.53}_{-0.58}$ \\
12 & 0625 & 52450.25 & 6.40$^{+0.23}_{-0.00}$ & 3.70$^{+  5.61}_{-2.47}$ & 0.038$^{+0.022}_{-0.027}$ & 132$^{+ 75}_{- 95}$ & 3.06$^{+0.48}_{-0.44}$ \\ 
13 & 0626 & 52451.57 & 6.40$^{+0.57}_{-0.00}$ & 3.04$^{+396.96}_{-1.81}$ & 0.017$^{+0.022}_{-0.017}$ &  68$^{+ 88}_{- 68}$ & 1.37$^{+0.48}_{-0.39}$ \\ 
14 & 0627 & 52452.69 & 6.40$^{+0.57}_{-0.00}$ & 1.24$^{+398.77}_{-0.00}$ & 0.020$^{+0.020}_{-0.020}$ &  81$^{+ 81}_{- 81}$ & 1.64$^{+0.43}_{-0.38}$ \\ 
15 & 0628 & 52453.50 & 6.78$^{+0.19}_{-0.25}$ & 3.04$^{+ 20.65}_{-1.81}$ & 0.036$^{+0.015}_{-0.016}$ & 208$^{+ 89}_{- 92}$ & 3.08$^{+0.35}_{-0.36}$ \\ 
16 & 0628 & 52453.55 & 6.71$^{+0.26}_{-0.31}$ & 3.37$^{+396.63}_{-2.14}$ & 0.026$^{+0.017}_{-0.023}$ & 146$^{+ 94}_{-128}$ & 2.19$^{+0.32}_{-0.27}$ \\ 
17 & 0629 & 52454.60 & 6.72$^{+0.25}_{-0.32}$ & 3.04$^{+396.96}_{-1.81}$ & 0.023$^{+0.015}_{-0.019}$ & 149$^{+ 99}_{-124}$ & 1.98$^{+0.31}_{-0.34}$ \\ 
18 & 0630 & 52455.13 & 6.82$^{+0.15}_{-0.42}$ & 3.70$^{+396.30}_{-2.47}$ & 0.022$^{+0.013}_{-0.020}$ & 183$^{+109}_{-167}$ & 1.93$^{+0.21}_{-0.19}$ \\ 
19 & 0701 & 52456.77 & 6.75$^{+0.22}_{-0.35}$ & 3.37$^{+396.63}_{-2.14}$ & 0.018$^{+0.011}_{-0.014}$ & 173$^{+105}_{-130}$ & 1.57$^{+0.23}_{-0.22}$ \\ 
20 & 0702 & 52457.83 & 6.78$^{+0.18}_{-0.33}$ & 3.04$^{+  8.34}_{-1.81}$ & 0.031$^{+0.011}_{-0.014}$ & 247$^{+ 91}_{-117}$ & 2.62$^{+0.22}_{-0.23}$ \\ 
21 & 0703 & 52458.44 & 6.72$^{+0.22}_{-0.32}$ & 3.21$^{+ 10.53}_{-1.97}$ & 0.028$^{+0.011}_{-0.015}$ & 233$^{+ 89}_{-121}$ & 2.40$^{+0.27}_{-0.22}$ \\ 
22 & 0704 & 52459.08 & 6.69$^{+0.22}_{-0.29}$ & 3.37$^{+ 12.59}_{-2.14}$ & 0.034$^{+0.013}_{-0.021}$ & 229$^{+ 87}_{-144}$ & 2.86$^{+0.21}_{-0.30}$ \\ 
23 & 0705 & 52460.49 & 6.79$^{+0.17}_{-0.30}$ & 3.37$^{+  3.65}_{-2.14}$ & 0.037$^{+0.010}_{-0.014}$ & 320$^{+ 85}_{-118}$ & 3.16$^{+0.19}_{-0.23}$ \\ 
24 & 0706 & 52461.41 & 6.76$^{+0.21}_{-0.36}$ & 3.04$^{+  9.07}_{-1.81}$ & 0.028$^{+0.011}_{-0.016}$ & 251$^{+100}_{-146}$ & 2.38$^{+0.20}_{-0.20}$ \\ 
25 & 0707 & 52462.57 & 6.75$^{+0.17}_{-0.20}$ & 3.37$^{+  2.88}_{-2.14}$ & 0.024$^{+0.006}_{-0.006}$ & 340$^{+ 84}_{- 91}$ & 2.01$^{+0.15}_{-0.17}$ \\ 
26 & 0708 & 52463.67 & 6.82$^{+0.14}_{-0.19}$ & 3.37$^{+  2.43}_{-2.14}$ & 0.020$^{+0.004}_{-0.005}$ & 389$^{+ 84}_{- 95}$ & 1.71$^{+0.09}_{-0.10}$ \\ 
27 & 0709 & 52464.45 & 6.83$^{+0.14}_{-0.43}$ & 3.54$^{+ 26.14}_{-2.30}$ & 0.006$^{+0.003}_{-0.003}$ & 245$^{+109}_{-140}$ & 0.52$^{+0.08}_{-0.08}$ \\ 
28 & 0710 & 52465.08 & 6.86$^{+0.11}_{-0.33}$ & 1.24$^{+ 11.99}_{-0.00}$ & 0.008$^{+0.002}_{-0.004}$ & 384$^{+113}_{-197}$ & 0.71$^{+0.07}_{-0.07}$ \\ 
29 & 0711 & 52466.34 & 6.81$^{+0.16}_{-0.20}$ & 3.37$^{+  1.68}_{-2.14}$ & 0.008$^{+0.002}_{-0.002}$ & 398$^{+ 95}_{-100}$ & 0.68$^{+0.05}_{-0.06}$ \\ 
30 & 0712 & 52467.20 & 6.67$^{+0.30}_{-0.27}$ & 3.70$^{+396.30}_{-2.47}$ & 0.005$^{+0.003}_{-0.005}$ & 232$^{+177}_{-232}$ & 0.38$^{+0.06}_{-0.08}$ \\ 
31 & 0713 & 52468.66 & 6.89$^{+0.08}_{-0.31}$ & 3.37$^{+  4.03}_{-2.14}$ & 0.005$^{+0.001}_{-0.002}$ & 468$^{+112}_{-202}$ & 0.42$^{+0.03}_{-0.05}$ \\ 
32 & 0713 & 52468.73 & 6.61$^{+0.30}_{-0.21}$ & 3.21$^{+ 21.06}_{-1.97}$ & 0.004$^{+0.002}_{-0.002}$ & 277$^{+136}_{-179}$ & 0.30$^{+0.05}_{-0.04}$ \\ 
33 & 0714 & 52469.24 & 6.85$^{+0.12}_{-0.24}$ & 3.04$^{+  7.25}_{-1.81}$ & 0.005$^{+0.001}_{-0.002}$ & 508$^{+134}_{-206}$ & 0.42$^{+0.04}_{-0.05}$ \\ 
34 & 0714 & 52469.31 & 6.67$^{+0.19}_{-0.27}$ & 3.37$^{+ 10.48}_{-2.14}$ & 0.004$^{+0.001}_{-0.002}$ & 395$^{+121}_{-200}$ & 0.37$^{+0.04}_{-0.03}$ \\ 
35 & 0715 & 52470.24 & 6.59$^{+0.38}_{-0.19}$ & 3.37$^{+396.63}_{-2.14}$ & 0.004$^{+0.001}_{-0.004}$ & 363$^{+141}_{-363}$ & 0.31$^{+0.05}_{-0.05}$ \\ 
36 & 0715 & 52470.30 & 6.40$^{+0.20}_{-0.00}$ & 1.24$^{+ 20.20}_{-0.00}$ & 0.002$^{+0.002}_{-0.002}$ & 153$^{+120}_{-149}$ & 0.16$^{+0.08}_{-0.04}$ \\ 
37 & 0716 & 52470.98 & 6.40$^{+0.57}_{-0.00}$ & 3.87$^{+396.13}_{-2.63}$ & 0.001$^{+0.002}_{-0.001}$ & 125$^{+218}_{-125}$ & 0.09$^{+0.08}_{-0.07}$ \\ 
38 & 0716 & 52471.02 & 6.84$^{+0.13}_{-0.44}$ & 1.24$^{+398.77}_{-0.00}$ & 0.001$^{+0.002}_{-0.001}$ & 153$^{+319}_{-153}$ & 0.09$^{+0.05}_{-0.05}$ \\ 
39 & 0717 & 52472.62 & 6.40$^{+0.27}_{-0.00}$ & 3.38$^{+ 10.53}_{-2.14}$ & 0.002$^{+0.002}_{-0.001}$ & 227$^{+292}_{-179}$ & 0.14$^{+0.04}_{-0.04}$ \\ 
40 & 0717 & 52472.69 & 6.56$^{+0.41}_{-0.16}$ & 3.70$^{+ 42.04}_{-2.47}$ & 0.002$^{+0.002}_{-0.002}$ & 274$^{+224}_{-274}$ & 0.16$^{+0.04}_{-0.04}$ \\ 
41 & 0718 & 52473.20 & 6.57$^{+0.14}_{-0.17}$ & 3.37$^{+ 12.63}_{-2.14}$ & 0.002$^{+0.002}_{-0.001}$ & 227$^{+181}_{-112}$ & 0.16$^{+0.04}_{-0.04}$ \\ 
42 & 0718 & 52473.27 & 6.40$^{+0.31}_{-0.00}$ & 1.24$^{+ 37.22}_{-0.00}$ & 0.002$^{+0.001}_{-0.002}$ & 202$^{+156}_{-178}$ & 0.14$^{+0.03}_{-0.04}$ \\ 
43 & 0719 & 52474.18 & 6.40$^{+0.20}_{-0.00}$ & 3.37$^{+  2.38}_{-2.14}$ & 0.003$^{+0.002}_{-0.002}$ & 313$^{+203}_{-210}$ & 0.28$^{+0.06}_{-0.05}$ \\ 
44 & 0720 & 52474.94 & 6.89$^{+0.08}_{-0.24}$ & 3.04$^{+  1.72}_{-1.81}$ & 0.004$^{+0.003}_{-0.001}$ & 442$^{+293}_{-137}$ & 0.36$^{+0.04}_{-0.05}$ \\ 
45 & 0721 & 52476.02 & 6.40$^{+0.12}_{-0.00}$ & 3.04$^{+  1.63}_{-1.81}$ & 0.004$^{+0.003}_{-0.001}$ & 289$^{+199}_{- 47}$ & 0.36$^{+0.05}_{-0.05}$ \\
46 & 0721 & 52476.08 & 6.43$^{+0.17}_{-0.04}$ & 3.04$^{+  3.19}_{-1.81}$ & 0.005$^{+0.002}_{-0.002}$ & 332$^{+162}_{-127}$ & 0.38$^{+0.04}_{-0.06}$ \\
47 & 0721 & 52476.15 & 6.40$^{+0.15}_{-0.00}$ & 1.40$^{+  4.65}_{-0.16}$ & 0.003$^{+0.001}_{-0.001}$ & 194$^{+ 76}_{- 63}$ & 0.23$^{+0.04}_{-0.04}$ \\
48 & 0722 & 52477.23 & 6.40$^{+0.10}_{-0.00}$ & 3.04$^{+  2.15}_{-1.81}$ & 0.003$^{+0.003}_{-0.001}$ & 228$^{+203}_{- 82}$ & 0.23$^{+0.06}_{-0.04}$ \\ 
49 & 0722 & 52477.30 & 6.40$^{+0.10}_{-0.00}$ & 3.04$^{+  1.69}_{-1.81}$ & 0.003$^{+0.002}_{-0.001}$ & 281$^{+159}_{-111}$ & 0.27$^{+0.03}_{-0.03}$ \\
\enddata 	    
\tablecomments{Output of spectral parameters from an input model
  consisting of a multi--color blackbody, power--law, Laor line
  emission, and a smeared absorption Fe edge.  $N_H$ = 4 \X 10$^{21}$ cm$^{-2}$
  was assumed  throughout.}
\tablenotetext{a}{Midpoint of observation.}
\tablenotetext{b}{`Laor'   model in XSPEC with the following fixed
  parameters: power--law=3, $i$=21\deg,  $R_{out}$=400$R_g$ ($R_g=GM/c^2$).}
\tablenotetext{c}{Inner radius in units of $R_g$.}
\tablenotetext{d}{Photons cm$^{-2}$ s$^{-1}$ in the line.}
\tablenotetext{e}{Unabsorbed flux in the 3--25~keV band.}
\end{deluxetable}
\clearpage

\begin{center}
\psfig{file=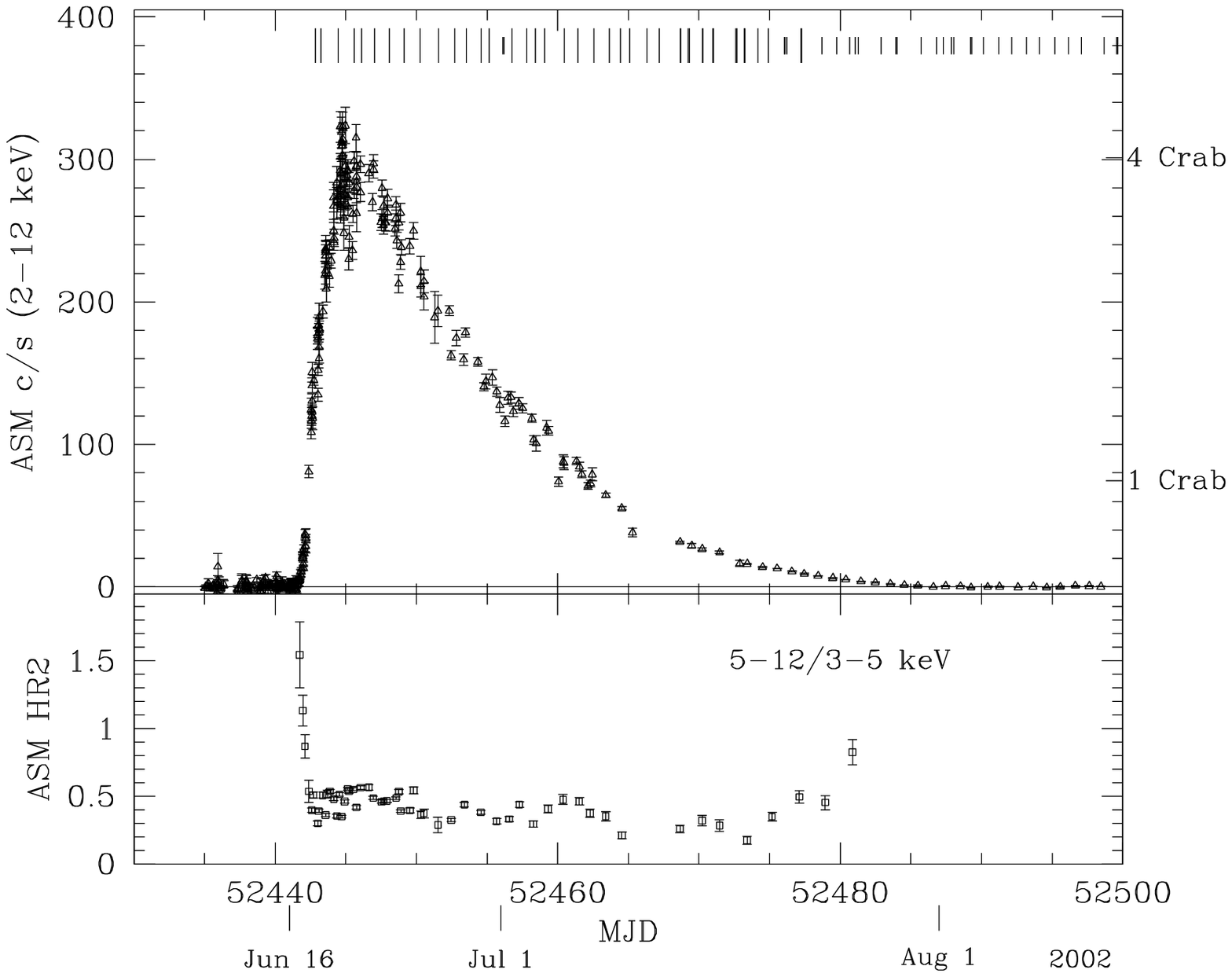, height=4.5in} 
\figcaption[asm.ps]{(Upper Panel) 2--12~keV ASM light curve and (Lower
Panel) hardness ratio (5--12~keV)/(3--5~keV) for \1543.  The vertical
lines at the top indicate the times of pointed RXTE observations; the
larger tickmarks indicate observations analyzed in this paper.}
\label{fig:asm}
\end{center}

\begin{center}
\psfig{file=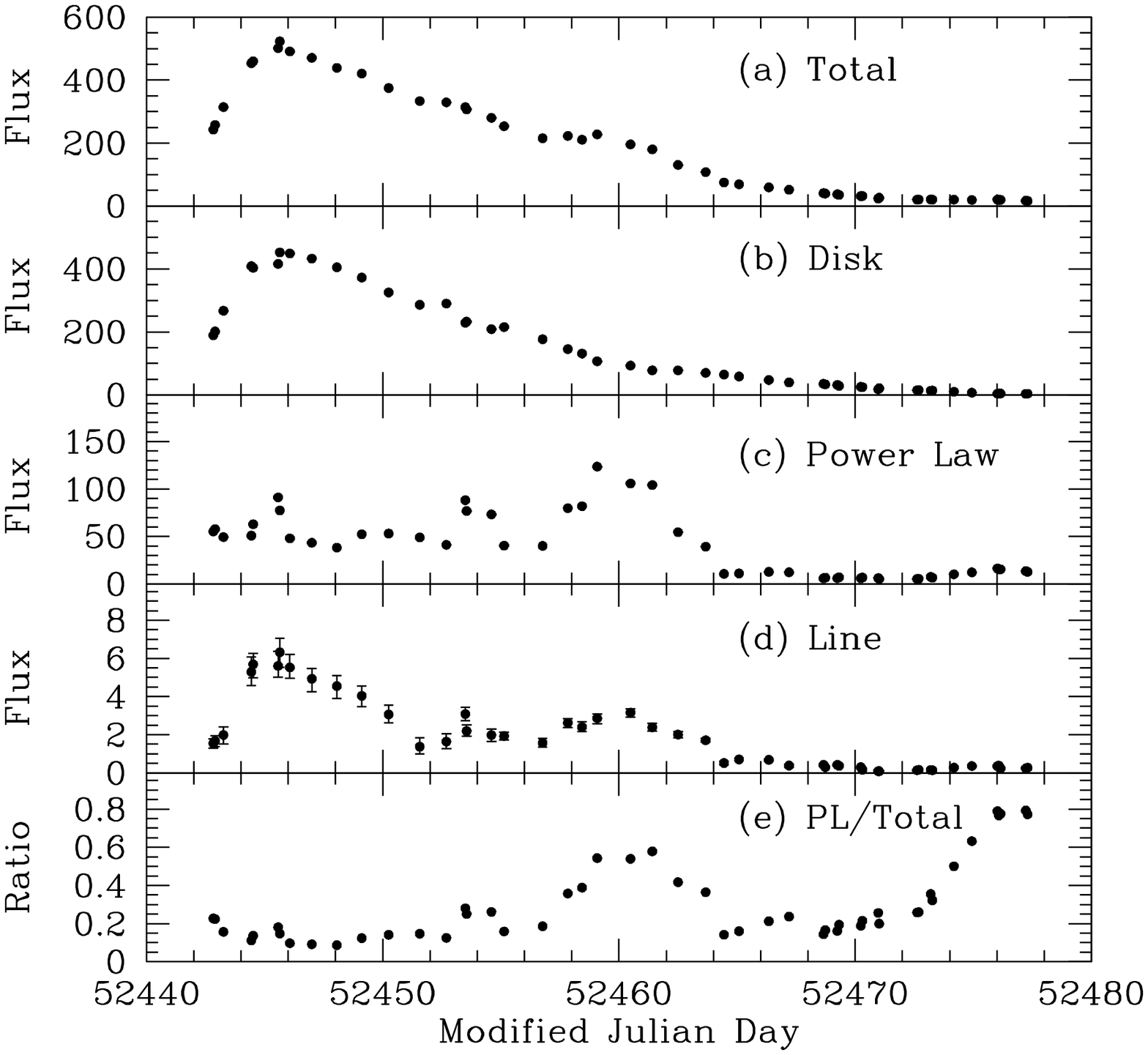, height=4in} 
\figcaption[flux.ps]{(a) Total flux, (b) disk flux, (c) power--law
flux, (d) line flux, and (e) ratio of the power--law flux to the total
flux for PCA observations of \1543 in the 3--25~keV band.  All fluxes
are given in units of 10$^{-10}$ \ergcm2s.  When error bars are not
visible, it is because they are smaller than the plotting
symbol. \label{fig:flux}}
\end{center}

\begin{center}
\psfig{file=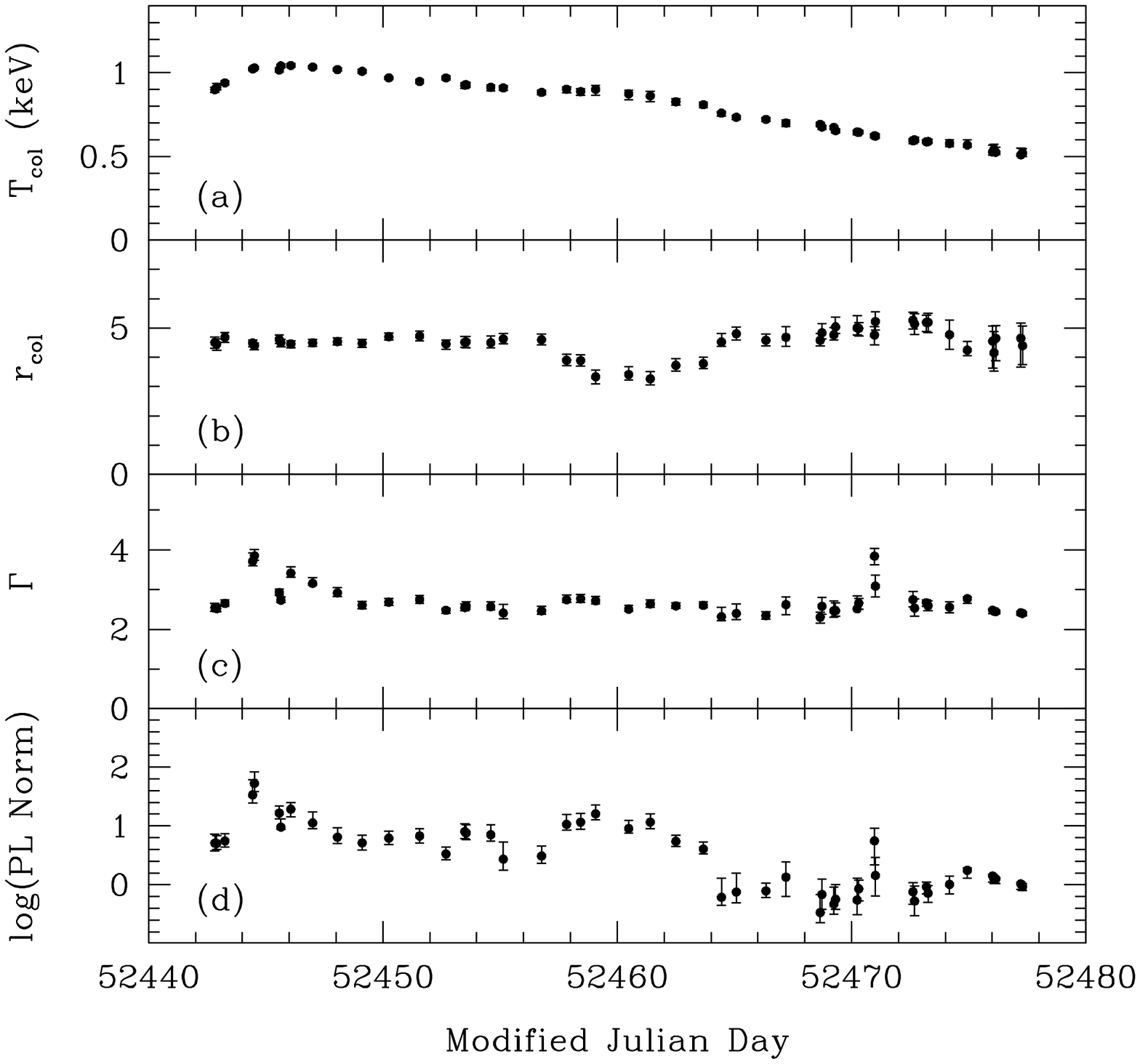, height=4in} 
\figcaption[cont.ps]{Spectral parameters for PCA observations of
\1543.  The plot shows (a) the temperature at the inner disk radius in
keV, (b) the inner disk radius in units of gravitational radii, (c)
the power--law photon index $\Gamma$, and (d) the power--law
normalization in units of photons keV$^{-1}$ cm$^{-2}$ s$^{-1}$ at 1
keV, plotted on a semi--log scale. When error bars are not visible, it
is because they are smaller than the plotting symbol.
\label{fig:cont}}
\end{center}

\begin{center}
\psfig{file=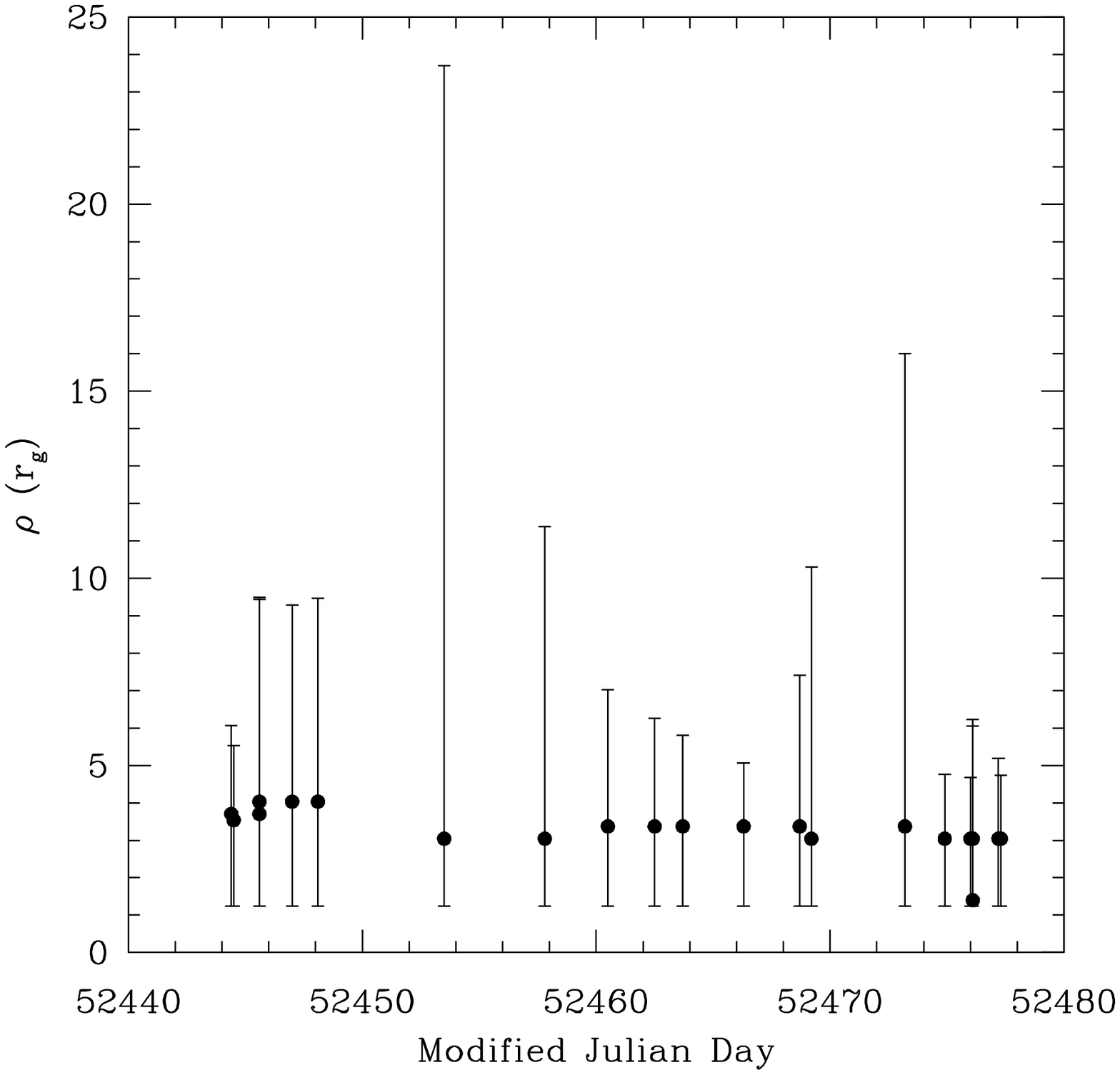, height=4in} 
\figcaption[rho.ps]{ Evolution of $\rho$.  The inner disk radius in
  units of gravitational radii as measured by the Laor line model is
  plotted against time.  Only those observations for which the
  equivalent width is greater than two times the minus side error in
  equivalent width are plotted.  The error bars are systematically
  large due to the poor resolution of {\it RXTE}, though the data
  points appear relatively stable.  The lower bound on each error bar
  is $\rho = 1.235$ because that is the radius of the innermost stable
  circular orbit for a prograde black hole with dimensionless angular
  momentum $a_{*} = 0.998$ (Thorne 1974). \label{fig:rho}}
\end{center}

\begin{center}
\psfig{file=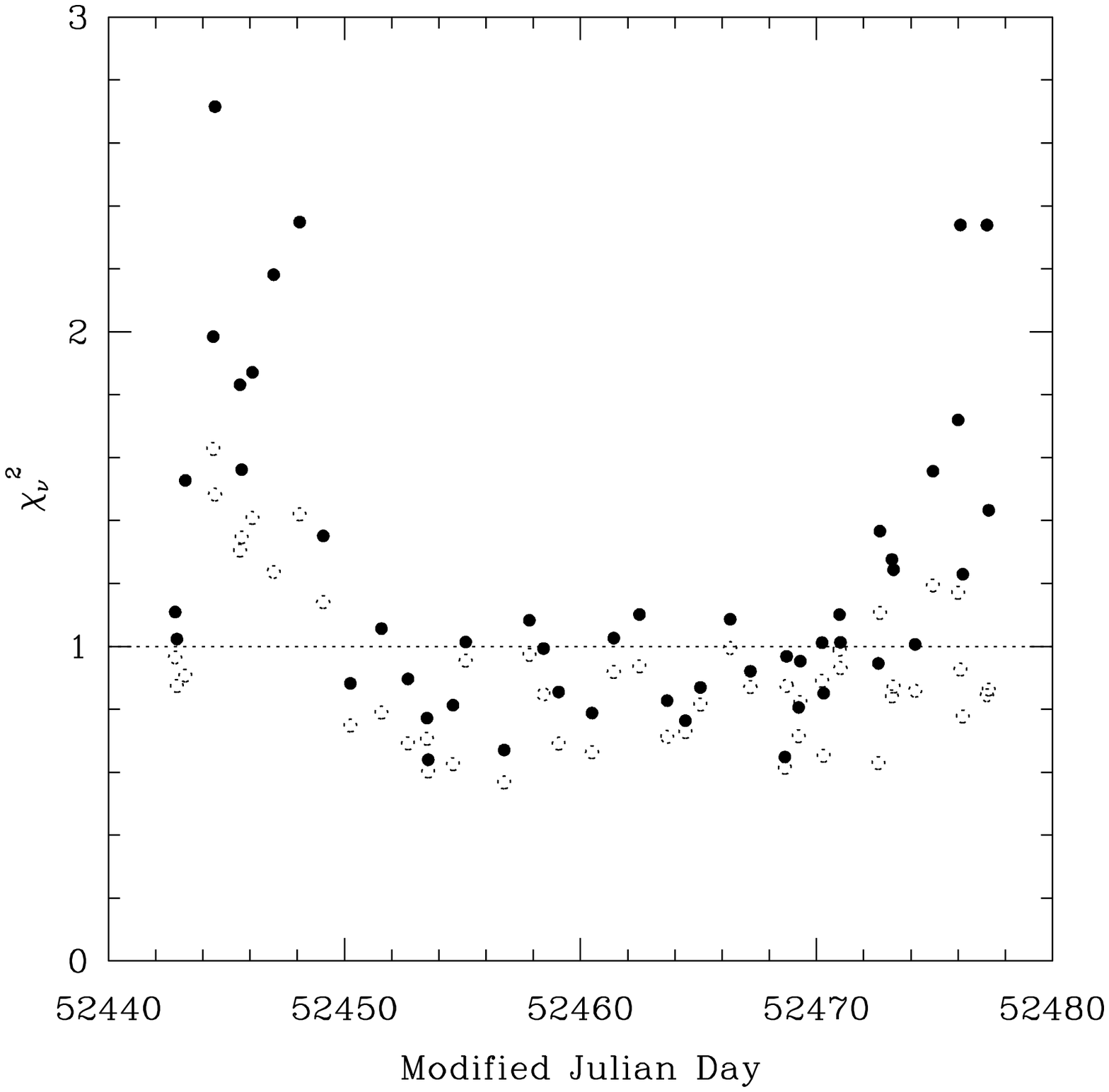, height=4in} 
\figcaption[chi.ps]{Comparison between the Laor model and the Gaussian
model for a fit of the Fe line component.  The solid circles show the
\redchi2 values for the Gaussian model.  The open circles show the
\redchi2 values for the Laor model.  Both models also include
interstellar absorption, disk blackbody, power--law, and smeared edge
components. \label{fig:lvsg}}
\end{center}

\begin{figure}[htbp]
\begin{center}
$
\begin{array}{cc}
\includegraphics[width=3.0in]{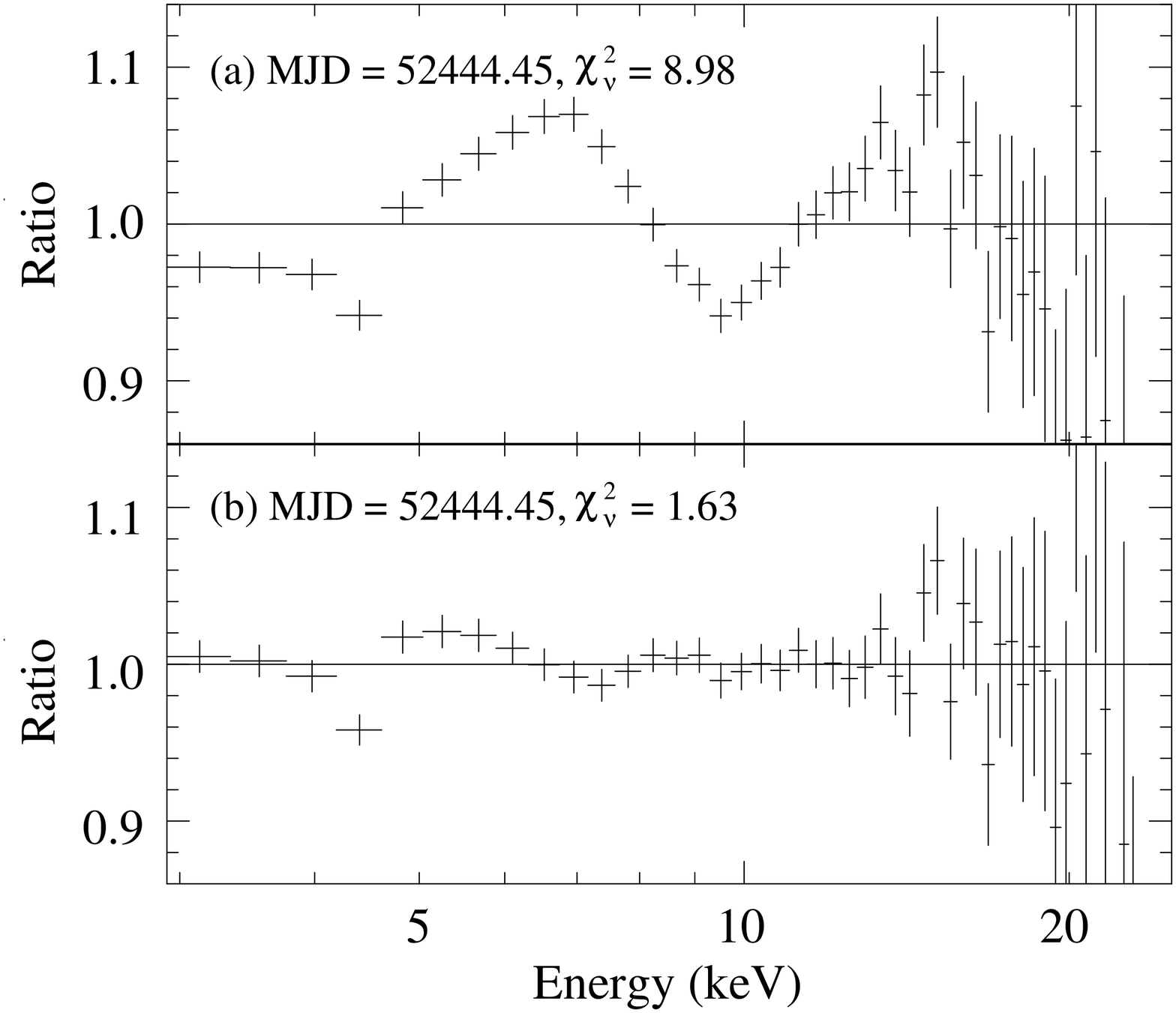} \\
\includegraphics[width=3.0in]{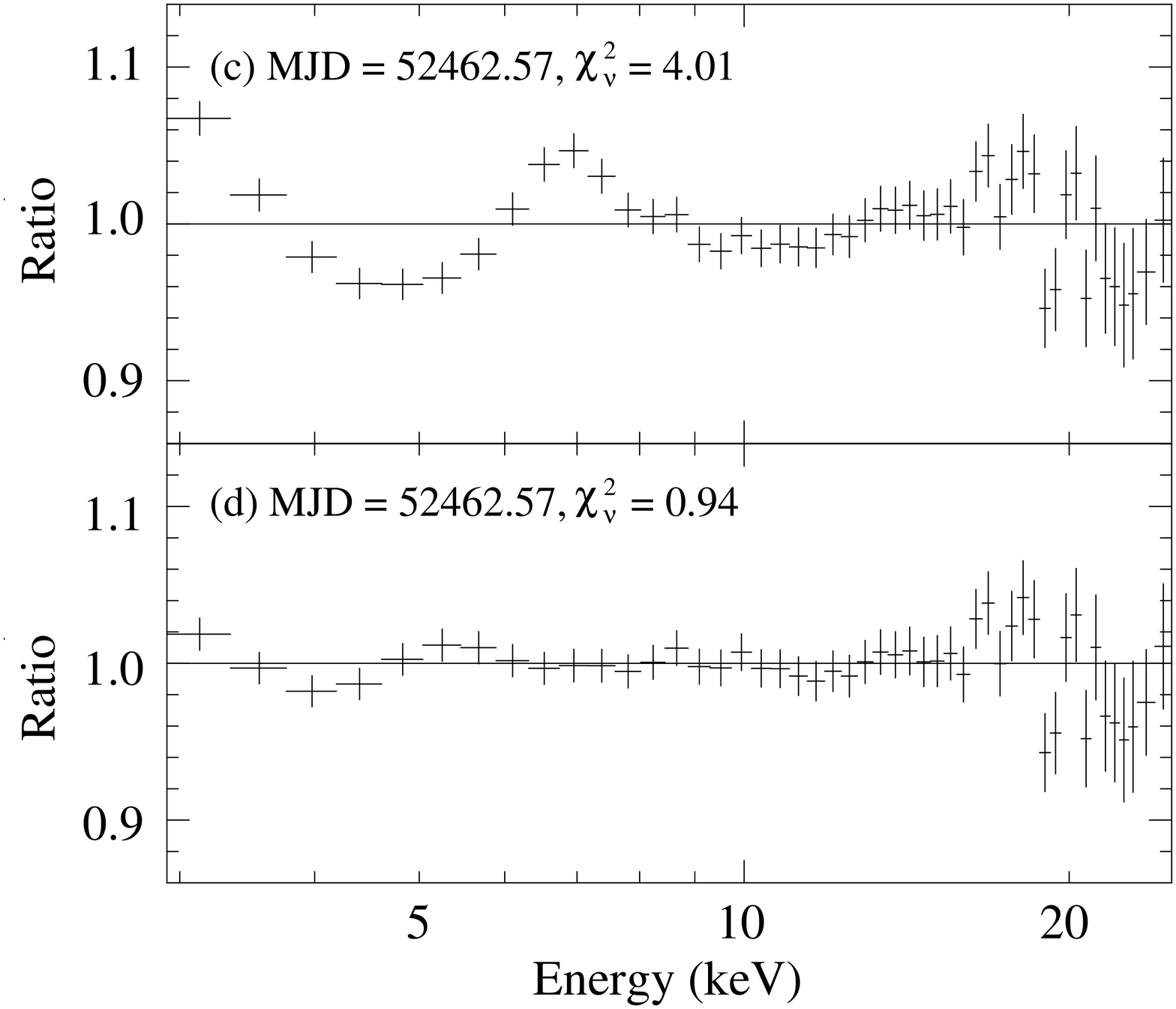} \\
\end{array} 
$
\caption{Sample PCA residuals.  Note in (a) the appearance of a strong,
  asymmetric Fe \Kalpha line.  Plots (a) and (c) show fits to a
  model consisting of a multi--color disk blackbody, power--law, and an
  interstellar absorption fixed at $N_H$ = 4 \X 10$^{21}$ cm$^{-2}$.
  Plots (b) and (d) show the fits after the inclusion of the Laor
  model and a smeared absorption edge.}\label{fig:xspec}
\end{center}
\end{figure}
\clearpage

\begin{center}
\psfig{file=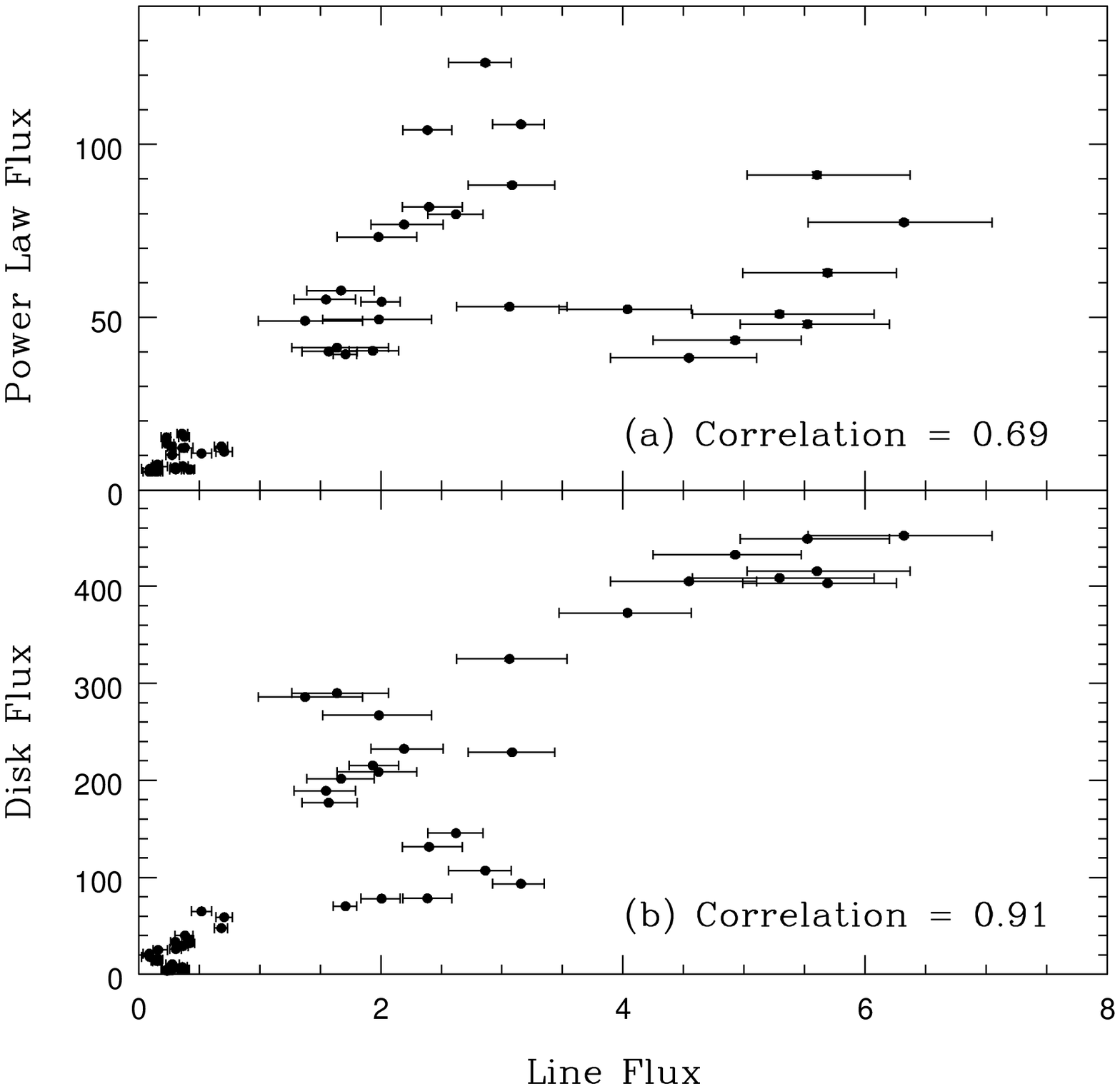, height=4in} 
\figcaption[corr.ps]{Correlations between line flux and other flux
parameters.  All fluxes given in units of 10$^{-10}$ \ergcm2s.  The
correlation coefficients were calculated using the linear Pearson
correlation test.  Errors are plotted for both the vertical and
horizontal directions.  When error bars are not visible, it is because
they are smaller than the plotting symbol.}\label{fig:corr}
\end{center}

\begin{center}
\psfig{file=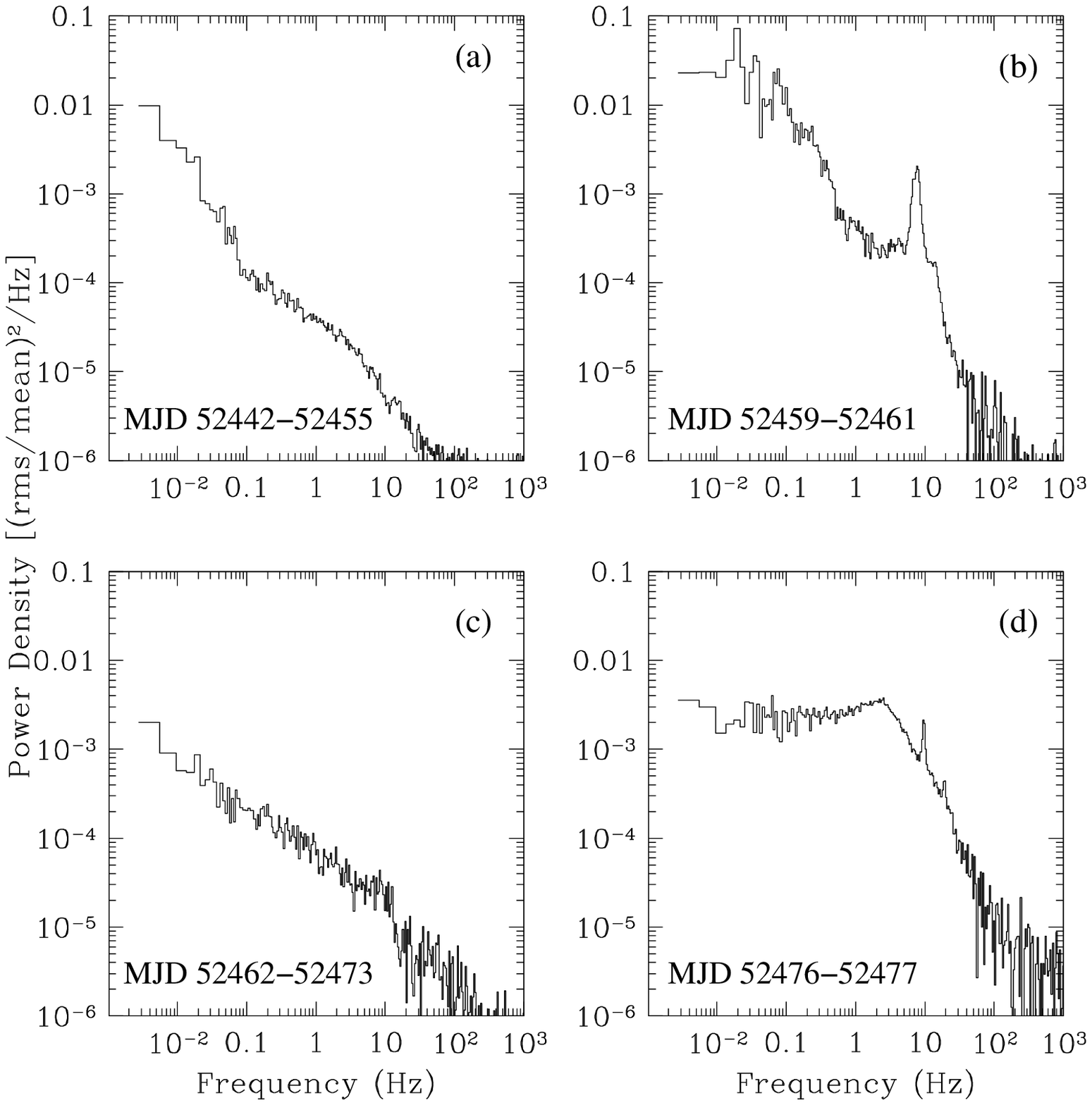, height=4in}
\figcaption[qpo.ps]{Average power density spectra for \1543.  PDS
  taken in the frequency range 4~mHz--4~kHz for data within 2--30~keV
  during four time intervals of the observation.  The gaps in the time
  intervals represent days when the source is in transition.  Note the
  presence of a strong QPO in panel (b).}\label{fig:qpo} 
\end{center}

\begin{center}
\psfig{file=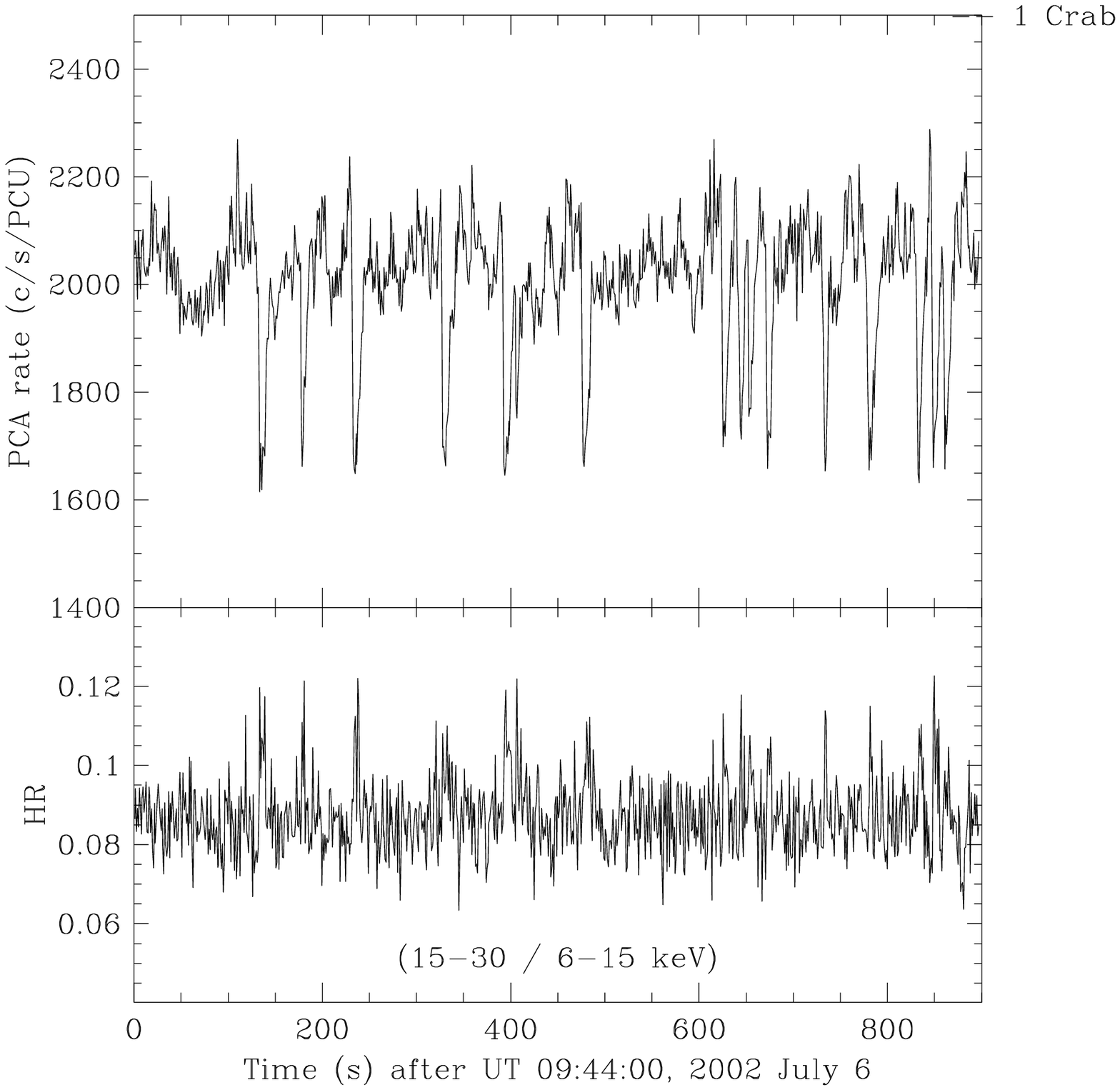, height=3.5in}
\figcaption[dip.ps]{Light curve of \1543 on MJD 52461.  The top panel
  shows the light curve with hard dips suggestive of an accretion
  instability.  The dips last 5--10 s, with a dip rate of roughly one
  per minute.  The PCA hardness ratio is plotted in the bottom panel.}
\label{fig:dip}
\end{center}

\begin{center}
\psfig{file=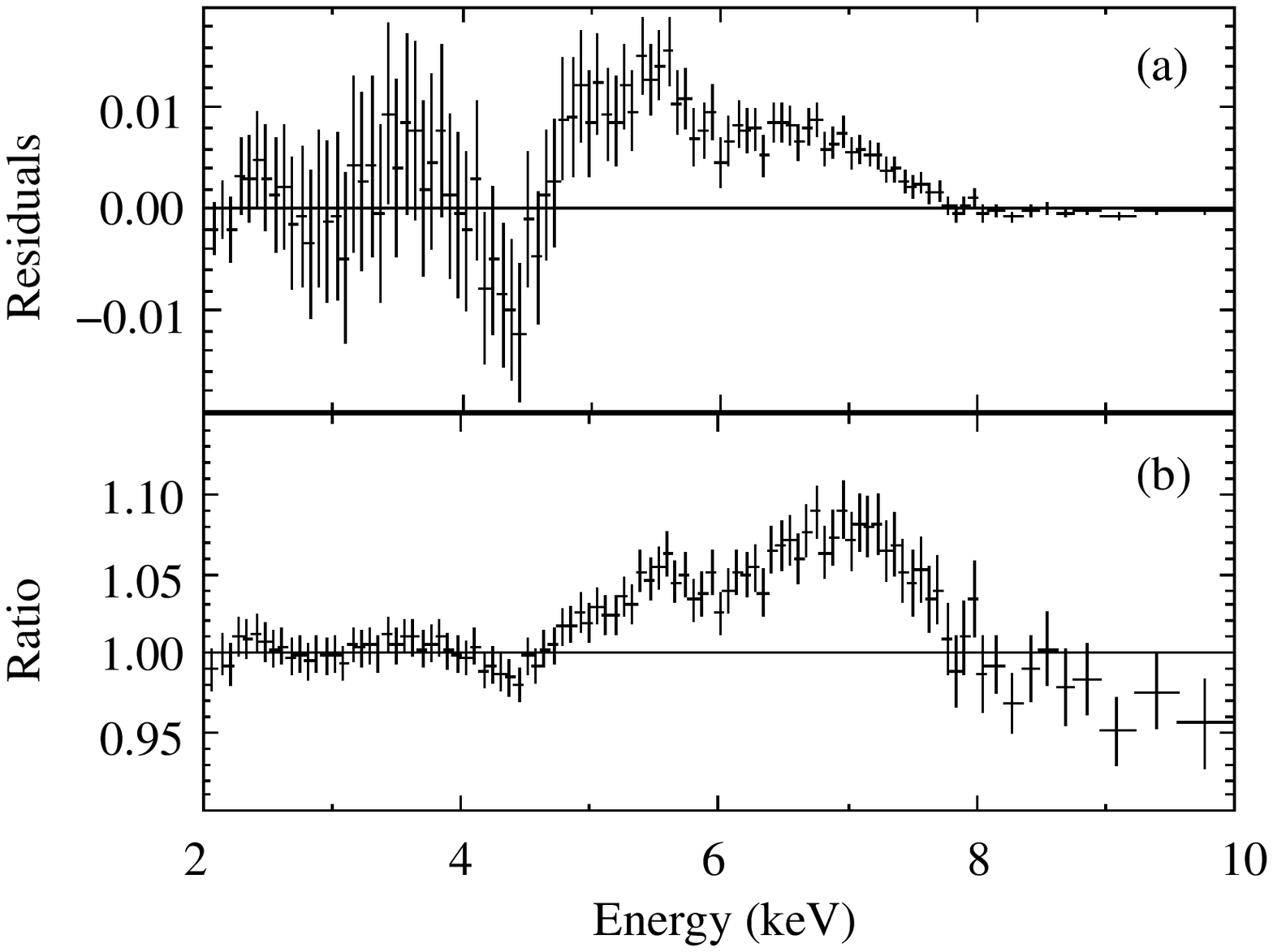, height=3in} 
\figcaption[resrat.eps]{Data from {\it EXOSAT}/GSPC taken in 1983 (see
van der Woerd et al.\ 1989).  The continuum is fit to a disk blackbody
and power--law model, and includes interstellar absorption.  The line
is fit to a Laor profile.  Panel (a) show the residuals in units of
normalized counts sec$^{-1}$ keV$^{-1}$.   Panel (b) shows the ratio
of the data to the model.  Note the clear double--peaked structure of
the line profile.}\label{fig:exosat}
\end{center}


\clearpage

\begin{thebibliography}{}
\expandafter\ifx\csname natexlab\endcsname\relax\def\natexlab#1{#1}\fi
\bibitem{arn96} Arnaud, K.~A. 1996, in ASP Conf. Ser. 101: Astronomical Data Analysis Software and Systems V., 17
\bibitem{bel00} Belloni, T., Klein-Wolt, M., Mendez, M., van der Klis, M., \& van  Paradijs, J. 2000, A\&A, 355, 271
\bibitem{che97} Chen, W., Shrader, C.~R., \& Livio, M.\ 1997, \apj, 491, 312
\bibitem{che92} Chevalier, C., Ilovaisky, S.~A.\ 1992, \iaucirc, 5520, 1
\bibitem{cze91} Czerny, B., Zbyszewska, M., \& Raine, D.~J.\ 1991, in Proc. Iron Line Diagnostics in X--ray Sources (Varenna 1990), ed. A.~Treves, G.~C.~Perola, \& L.~Stella, Lecture Notes in Physics (Berlin: Springer--Verlag), vol. 385, 226
\bibitem{dic90} Dickey, J.~M. \& Lockman, F.~J.\ 1990, \araa, 28, 215
\bibitem{die00} Dieters, S.~W., et al. 2000, \apj, 538, 307
\bibitem{ebi94} Ebisawa, K., et al.\ 1994, \pasj, 46, 375 
\bibitem{fab00} Fabian, A.~C., Iwasawa, K., Reynolds, C.~S. \& Young, A.~J.\ 2000, \pasp, 112, 1145
\bibitem{fab95} Fabian, A.~C., Nandra, K., Reynolds, C.~S., Brandt, W.~N., Otani, C., Tanaka, Y., Inoue, H., \& Iwasawa, K.\ 1995, \mnras, 277, L11
\bibitem{fen03} Fender, R.~P., 2003, to appear in "Compact Stellar X-ray Sources", eds. W.~H.~G.\ Lewin and M.\ van der Klis, Cambridge: Cambridge University Press
\bibitem{gar01} Garcia, M.~R., McClintock, J.~E., Narayan, R., Callanan, P., Barret, D., \& Murray, S.~S.\ 2001, \apjl, 553, L47
\bibitem{geo91} George, I.~M. \& Fabian, A.~C.\, 1991, \mnras, 249, 352
\bibitem{han00} Hannikainen, D., et al., 2000, ApSSS, 276, 45
\bibitem{han00a} Hannikainen, D.~C., Hunstead, R.~W., Campbell-Wilson, D., Wu, K., McKay, D.~J., Smits, D.~P., \& Sault, R.~J.\ 2000 \apj, 540, 521
\bibitem{har92} Harmon, B.~A., Wilson, R.~B., Finger, M.~H., Paciesas, W.~S., Rubin, B.~C., \& Fishman, G.~J.\ 1992, \iaucirc, 5504, 1
\bibitem{har95} Harmon, B.~A., et al.\ 1995, \nat, 374, 703
\bibitem{hje95} Hjellming, R.~M., \& Rupen, M.~P.\ 1995, \nat, 375, 464
\bibitem{ibr03} Ibrahim, A.~I., Swank, J.~H., \& Parke, W.\ 2003, \apjl, 584, L17
\bibitem{iwa96} Iwasawa, K., et al.\ 1996, \mnras, 282, 1038
\bibitem{kal02} Kalemci, E., Tomsick, J., Rothschild, R., Corbel, S., Kaaret, P., \& McClintock, J.\ 2002, ATEL, 103
\bibitem{kit84} Kitamoto, S., Miyamoto, S., Tsunemi, H., Makishima, K., \& Nakagawa, M.\ 1984, \pasj, 36, 799
\bibitem{lao91} Laor, A.\ 1991, \apj, 376, 90
\bibitem{mak86} Makishima, K., Maejima, Y., Mitsuda, K., Bradt, H.~V.,   Remillard, R.~A., Tuohy, I.~R., Hoshi, R., \& Nakagawa, M.\ 1986, \apj, 308, 635
\bibitem{mar01} Markoff, S., Falcke, H., \& Fender, R.\ 2001, \aap, 372, L25
\bibitem{mar03} Markoff, S., Nowak, M., Corbel, S., Fender, R., \& Falcke, H.\ 2003, \aap, 397, 645
\bibitem{mat72} Matilsky, T.~A., Giacconi, R., Gursky, H., Kellogg, E.~M., \& Tananbaum, H.~D.\, 1972, \apjl, 174, L53
\bibitem{mcc03} McClintock, J.~E., \& Remillard, R.~A.\ 2003, to appear in Compact Stellar X--ray Sources, ed. W.~H.~G.~Lewin \& M.~van der Klis, Cambridge: Cambridge Univ. Press, preprint (astro-ph/0306213)
\bibitem{mer00} Merloni, A., Fabian, A.~C., \& Ross, R.~R.\ 2000, \mnras, 313, 193
\bibitem{mil02} Miller, J.~M., et al.\ 2002, \apj, 578, 348
\bibitem{mil01} Miller, J.~M., Fox, D.~W., Di Matteo, T., Wijnands, R., Belloni, T., Pooley, D., Kouveliotou, C., \& Lewin, W.~H.~G.\ 2001, \apj, 546, 1055
\bibitem{mit84} Mitsuda, K., et al.\ 1984, \pasj, 36, 741
\bibitem{oro03} Orosz, J.~A.\ 2003, in preparation
\bibitem{oro98} Orosz, J.~A. and Jain, R.~K., Bailyn, C.~D., McClintock, J.~E., \& Remillard, R.~A.\ 1998, \apj, 499, 375
\bibitem{ped83} Pedersen, H.\ 1983, The Messenger, 34, 21
\bibitem{pet01} Petrucci, P.~O., Merloni, A., Fabian, A., Haardt, F., \& Gallo, E.\ 2001, \mnras, 328, 501
\bibitem{rem99} Remillard, R.~A., Morgan, E.~H., McClintock, J.~E., Bailyn, C.~D., \& Orosz, J.~A. 1999, \apj, 522, 397
\bibitem{rem02} Remillard, R.~A., Sobczak, G.~J.,, Muno, M.~ P., \& McClintock, J.~E.\ 2002, \apj, 564, 962
\bibitem{rey03} Reynolds, C.~S., \& Nowak, M.~A.\, 2003, Phys. Rept., 377, 389
\bibitem{rho74} Rhoades, C.~E., \& Ruffini, R.\, 1974, \prl, 32, 324
\bibitem{ros99} Ross, R.~R., Fabian, A.~C., Young, A.~J.\ 1999, \mnras, 306, 461
\bibitem{sha73} Shakura, N.~I., \& Sunyaev, R.~A.\ 1973, \aap, 24, 337
\bibitem{sha83} Shapiro, S.~L., \& Teukolsky, S.~A.\ 1983, Black Holes, White Dwarfs, and Neutron Stars: The Physics of Compact Objects (New York: Wiley)
\bibitem{shi95} Shimura, T., \& Takahara, F.\ 1995, \apj, 445, 780
\bibitem{sob00} Sobczak, G.~J., McClintock, J.~E., Remillard, R.~A., Cui, W., Levine, A.~M., Morgan, E.~H., Orosz, J.~A., \& Bailyn, C.~D.\ 2000, \apj, 544, 993
\bibitem{tan95} Tanaka, Y., \& Lewin, W.~H.~G.\ 1995, X--ray Binaries, ed. W.~H.~G.~Lewin, J.~van~Paradijs, \& E.~P.~J.~van~den~Heuvel (Cambridge: Cambridge Univ. Press) 
\bibitem{tit02} Titarchuk, L., \& Shrader, C.~R.\ 2002, \apj, 567, 1057 
\bibitem{tho74} Thorne, K.~S., 1974, \apj, 191, 507
\bibitem{tom98} Tomsick, J.~A., Lapshov, I., \& Kaaret, P. 1998, \apj, 494, 747
\bibitem{van89} van der Woerd, H., White, N.~E., \& Kahn, S.~M.\ 1989, \apj, 344, 320
\bibitem{van95} van Paradijs, J., \& McClintock, J.~E.\ 1995, Optical and Ultraviolet Observations of X--ray Binaries, X--ray Binaries, eds.~W.H.G.~Lewin, J.~van Paradijs, and E.P.J.~van den Heuvel (Cambridge: Cambridge Univ.~Press), 58  
\bibitem{zdz03} Zdziarski, A.~A., Lubi{\' n}ski, P., Gilfanov, M., \& Revnivtsev, M.\ 2003, \mnras, 342, 355
\bibitem{zha97} Zhang, S.~N., Cui, W., \& Chen, W.\ 1997, \apjl, 482, L155


\end{thebibliography}
\end{document}